%% file: COS3018_main.tex
\documentclass[fleqn,usenatbib]{mnras}

\usepackage[T1]{fontenc}

\DeclareRobustCommand{\VAN}[3]{#2}
\let\VANthebibliography\thebibliography
\def\thebibliography{\DeclareRobustCommand{\VAN}[3]{##3}\VANthebibliography}

\usepackage{graphicx}	
\usepackage{amsmath}	
\usepackage{comment}
\usepackage{txfonts}
\usepackage{hyperref}
\usepackage{textgreek}
\usepackage{xargs}
\usepackage{xspace}
\usepackage[dvipsnames,hyperref]{xcolor}

\usepackage{hyperref}
\hypersetup{
    colorlinks=true,
    linkcolor=Cerulean,
    citecolor=NavyBlue,
    filecolor=magenta,
    urlcolor=Violet,
    pdftitle= {COS3018-JWST+ALMA},
    pdfpagemode=FullScreen,
    }

\newcommand{\Msun}{\ensuremath{\mathrm{M}_\odot}\xspace}
\newcommand{\jwst}{\textit{JWST}\xspace}

\newcommand{\Halpha}{\text{H\textalpha}\xspace}
\newcommand{\Hbeta}{\text{H\textbeta}\xspace}
\newcommandx{\permittedEL}[6][1=O,2=III,3=,4=,5=,6=]{\text{{#1}\,{\sc {#2}}{#3}{#4}{#5}{#6}}\xspace}
\newcommandx{\semiforbiddenEL}[6][1=O,2=III,3=,4=,5=,6=]{\text{{#1}\,{\sc {#2}}]{#3}{#4}{#5}{#6}}\xspace}
\newcommandx{\forbiddenEL}[6][1=O,2=III,3=,4=,5=,6=]{\text{[{#1}\,{\sc{#2}}]{#3}{#4}{#5}{#6}}\xspace}
\newcommand{\kms}{km s$^{-1}$}

\newcommand{\HeIIL}[1][1=1640]{\permittedEL[He][ii][\textlambda][#1]}
\newcommand{\OII}{\forbiddenEL[O][ii]}
\newcommandx{\OIIL}[1][1=3728]{\forbiddenEL[O][ii][\textlambda][#1]}
\newcommand{\OIIall}{\forbiddenEL[O][ii][\textlambda][\textlambda][3727,][3729]}
\newcommand{\OIII}{\forbiddenEL[O][iii]}
\newcommandx{\OIIIL}[1][1=5007]{\forbiddenEL[O][iii][\textlambda][#1]}
\newcommand{\OIIIall}{\forbiddenEL[O][iii][\textlambda][\textlambda][5007,][4959]}
\newcommand{\NII}{\forbiddenEL[N][ii]}
\newcommandx{\NIIL}[1][1=6585]{\forbiddenEL[N][ii][\textlambda][#1]}
\newcommand{\NIIall}{\forbiddenEL[N][ii][\textlambda][\textlambda][6550,][6585]}

\newcommandx{\NeIIIL}[1][1=3869]{\forbiddenEL[Ne][iii][\textlambda][#1]}
\newcommand{\NeIIIall}{\forbiddenEL[Ne][iii][\textlambda][\textlambda][3869,][3968]}
\newcommand{\CIIIall}{\semiforbiddenEL[C][iii][\textlambda][\textlambda][1907,][1909]}
\newcommand{\CIVall}{\permittedEL[C][iv][\textlambda][\textlambda][1548,][1551]}

\newcommand{\CII}{\forbiddenEL[C][ii][\textlambda][158$\mu$m]}

\newcommand{\NeIV}{\forbiddenEL[Ne][iv]}
\newcommand{\NeV}{\forbiddenEL[Ne][v]}

\newcommand{\Te}{\ensuremath{T_\text{e}}\xspace}
\newcommand{\Tiii}{\ensuremath{t_3}\xspace}
\newcommand{\Tii}{\ensuremath{t_2}\xspace}

\newcommand{\target}{COS-3018\xspace}
\newcommand{\ergscm}{erg s$^{-1}$ cm$^{-2}$\xspace}
\newcommand{\JWST}{\textit{JWST}\xspace}

\title[COS-3018]{GA-NIFS: ISM properties and metal enrichment in a merger-driven starburst during the Epoch of Reionisation probed with JWST and ALMA}

\author[J. Scholtz]{\parbox[h]{\textwidth}{
J.\ Scholtz$,^{\! 1,2}$\thanks{E-mail: js2685@cam.ac.uk}
M. Curti,$^{3}$
F. D'Eugenio,$^{1,2}$
H. \"Ubler,$^{4,1,2}$
R. Maiolino,$^{1,2,5}$
C. Marconcini,$^{6,7}$
R. Smit,$^{8}$
M. Perna,$^{9}$
J. Witstok,$^{1,2}$
S, Arribas,$^{9}$
T. B\"oker,$^{10}$
A. J. Bunker,$^{11}$
S. Carniani,$^{12}$
S. Charlot,$^{13}$
G. Cresci,$^{6}$
P. G. P\'erez-Gonz\'alez,$^{9}$
I. Lamperti,$^{6}$
B. Rodr\'iguez Del Pino,$^{5}$
E. Parlanti,$^{12}$
G. Venturi$^{12}$
}\vspace{0.4cm}
\\
$^{1}$Kavli Institute for Cosmology, University of Cambridge, Madingley Road, Cambridge, 
CB3 0HA, UK\\
$^{2}$Cavendish Laboratory, University of Cambridge, 19 JJ Thomson Avenue, Cambridge CB3 0HE, UK\\
$^{3}$European Southern Observatory, Karl-Schwarzschild-Strasse 2, 85748 Garching, Germany\\
$^{4}$Max-Planck-Institut f\"ur extraterrestrische Physik (MPE), Gie{\ss}enbachstra{\ss}e 1, 85748 Garching, Germany\\
$^{5}$Department of Physics and Astronomy, University College London, Gower Street, London WC1E 6BT, UK\\
$^{6}$ INAF - Osservatorio Astrofisico di Arcetri, largo E. Fermi 5, 50127 Firenze, Italy\\
$^{7}$ 1.Dipartimento di Fisica e Astronomia, Università di Firenze, Via G. Sansone 1, 50019, Sesto F.no (Firenze), Italy\\
$^{8}$Astrophysics Research Institute, Liverpool John Moores University, 146 Brownlow Hill, Liverpool L3 5RF, UK \\
$^{9}$Centro de Astrobiolog\'ia (CAB), CSIC–INTA, Cra. de Ajalvir Km.~4, 28850- Torrej\'on de Ardoz, Madrid, Spain\\
$^{10}$European Space Agency, c/o STScI, 3700 San Martin Drive, Baltimore, MD 21218, USA\\
$^{11}$ University of Oxford, Department of Physics, Denys Wilkinson Building, Keble Road, Oxford OX13RH, United Kingdom\\
$^{12}$ Scuola Normale Superiore, Piazza dei Cavalieri 7, I-56126 Pisa, Italy\\
$^{13}$Sorbonne Universit\'e, CNRS, UMR 7095, Institut d'Astrophysique de Paris, 98 bis bd Arago, 75014 Paris, France\\
}

\date{Accepted XXX. Received YYY; in original form ZZZ}

\pubyear{2024}

\begin{document}
\label{firstpage}
\pagerange{\pageref{firstpage}--\pageref{lastpage}}
\maketitle

\begin{abstract}
We present deep \jwst/NIRSpec integral-field spectroscopy (IFS) and ALMA \CII observations of \target, a star-forming galaxy at z$\sim$6.85, as part of the GA-NIFS programme. Both G395H (R$\sim$ 2700) and PRISM (R$\sim$ 100) NIRSpec observations revealed that \target is comprised of three separate components detected in \OIIIL[5007], which we dub as Main, North and East, with stellar masses of 10$^{9.4 \pm 0.1}$, 10$^{9.2 \pm 0.07}$, 10$^{7.7 \pm 0.15}$ \Msun. We detect \OIIIall, \OIIall and multiple Balmer lines in all three components together with \OIIIL[4363] in the Main and North components. This allows us to measure an ISM temperature of \Te=~1.27$\pm0.07\times 10^4$ and \Te=~1.6$\pm0.14\times 10^4$ K with densities of $n_{e}$~=~1250$\pm$250 and $n_{e}$~=~700$\pm$200 cm$^{-3}$, respectively. These deep observations allow us to measure an average metallicity of 12+log(O/H)=7.9--8.2 for the three components with the  T$_{e}$-method. We do not find any significant evidence of metallicity gradients between the components. Furthermore, we also detect \NIIL[6585], one of the highest redshift detections of this emission line. We find that in a small, metal-poor clump 0.2\arcsec\ west of the North component, N/O is elevated compared to other regions,  indicating that nitrogen enrichment originates from smaller substructures, possibly proto-globular clusters. \OIIIL[5007] kinematics show that this system is merging, which is probably driving the ongoing, luminous starburst. 

\end{abstract}




\section{Introduction} 

With the launch of James Webb Space Telescope (\JWST), we are now able to observe rest-frame optical and UV emission features, and hence probe the interstellar medium (ISM) of galaxies, up to redshift $\sim$ 14 \citep{Arribas23,Cameron23,Curtis-lake23, Harikane23,Larson23,Isobe23,Hsiao23,Robertson23, Abdurrouf24, Carniani24, Harikane24,Sanders24,Tacchella23,Tacchella24, Vikaeus24}.

Before the launch of \JWST, the main avenue to study the ISM properties of galaxies at the Epoch of Reionisation (EoR; z$>$6) was through \CII and \OIIIL[88$\mu$m] emission lines and dust continuum observed with mm/sub-mm facilities (mainly ALMA). These observations revealed the early emergence of rotating discs \citep[e.g.][]{Smit18, Neelman20, Rizzo20, Fraternali21, Lelli21, Rizzo21, Parlanti23_alm,Rowland24} and a fast production of dust \citep[e.g. ][]{Laporte17,Witstok22, Bouwens21}. However, the ISM studies were severely limited by the lack of access to rest-frame optical emission lines as well as the limited detectability of FIR lines by ALMA.\jwst has demonstrated its ability to spatially resolve the ISM  at very high-z, opening the opportunity to study not only the global properties but also the internal structure of early cosmic systems \citep[e.g.][]{Arribas23,Decarli24, DEugenio23ifs, Jones24, Lamperti24, Rodriguezdelpino24, Ubler24b}.

With its unmatched capabilities, NIRSpec (near infra-red spectrograph) on board \JWST has enabled rapid progress in the physical properties of galaxies at z$>$3 when it comes to their abundance \citep{Perez-Gonzalez23, harikane_uv_LF_2024, robertson_JOF_LF_2024,mcleod_LF_2024}, detection of active galactic nuclei (AGN, e.g. \citealt{furtak23,greene23, Harikane23, Kocevski23, Maiolino23gnz11,Maiolino23JADES, Matthee23, Scholtz23AGN,Perna23_dual,Onoue23,Ubler23, Ubler24, Ubler24b}), bursty star-formation histories (SFHs) \citep[e.g.][]{dressler23,Endsley23, Looser23b, Tacchella22, Clarke24}, discovery of compact galaxies with intense starbursts and/or nuclear activity enshrouded by significant amounts of warm dust \citep{Akins24, Casey24, Perez-Gonzalez24} and ISM conditions \citep[e.g.][]{Sanders23, Cameron23, Reddy23, Calabro24}. 

\JWST observations of high redshift galaxies have revealed metal-poor galaxies with intense star formation, releasing large amounts of ionising radiation resulting in high ionisation parameters \citep[][]{Hirschmann22, Curti24JADES, Curti23ERO,Tacchella23, Trump23, Simmonds24}. Furthermore, the access to deep observations of rest-frame optical and UV emission lines allows astronomers to investigate the abundances of different elements, revealing, in some cases, unexpected ionisation and chemical enrichment patterns \citep[e.g.][]{Bunker23gnz11, Maiolino23gnz11, Cameron23gnz11, isobe_density_jwst_2023, Cameron23_9422,Topping24, DEugenio23z12, Schaerer24z94,Ji2024_nitrogen,Ji_GNz11_2024}. \jwst/NIRSpec has also allowed for significant progress in galaxy kinematics, as we now have access to optical emission lines at z$>$3.5 \citep{Nelson23,Ubler24b, Jones24,Lamperti24, Rodriguezdelpino24, degraaff24}, allowing for a comparison of ionised and cold gas kinematics at high redshift \citep{Parlanti23,Lamperti24}.

In this paper, we present new observations of \target from the Galaxy Assembly with NIRSpec Integral Field Spectroscopy (GA-NIFS)
Guaranteed Time Observations (GTO) programme (e.g. \citealt{Arribas23, Marshall23, Perna23b, Ubler23, Ji24, DEugenio23ifs, Jones24, Lamperti24, Rodriguezdelpino24, Perez-gonzalez-jekyll, Ubler24, Ubler24b}). This survey aims to investigate the spatially resolved stellar populations, ISM, outflow and kinematics properties of 55 quasars, AGN and star-forming galaxies (SFGs) in the redshift range of $z\sim2-11$ with NIRSpec IFU, utilising both the PRISM, medium and high spectral resolution observations. In this work, we present spatially resolved gas and stellar populations of \target using new \JWST/NIRSpec Integral Field Unit (IFU) high-resolution grating (R $\sim$ 2700) and low-resolution prism (R $\sim$ 100) observations as well as \JWST/NIRCam imaging.

\target, a star-forming galaxy at z=6.85, was first discovered by \citet{Smit14} by identifying objects with large equivalent width (EW) of \OIIIall+ \Hbeta based on \textit{HST} and \textit{Spitzer}/IRAC photometry (EW$_{\rm rest}>1200~\AA$).  \target has been intensively studied using both ground-based near-infrared (NIR) spectroscopy and ALMA. \citet{Laporte17} used VLT/X-shooter \citep{Vernet11} to detect \CIIIall emission, without any detection of Ly\textalpha\ or higher-ionisation lines like \HeIIL[1640] and \CIVall. ALMA observations of \CII emission lines revealed a velocity gradient in the system, suggesting an established cold rotating disc at these early epochs \citep{Smit18}, later confirmed by \citet{Parlanti23} as a turbulent disk with high-velocity dispersion. \citet{Vallini2020} studied the ISM conditions in \target using  \CIIIall, \CII and \OIIIL[88$\mu$m] observations, inferring a high metallicity of $\sim 0.4$ Z$_{\odot}$ and an ISM density of $\sim 500$ cm$^{-3}$. \citet{Witstok22} further analysed the ALMA dust continuum and integrated emission line properties to constrain the dust temperature of $\sim$30-40~K, resulting in a high dust mass of 2-26$\times 10^{7}$ \Msun. This dust mass measurement implies a dust-to-stellar mass ratio of 5 per cent, challenging theoretical models to create this much dust by such a high redshift. \target was observed by the PRIMER programme covering the target with NIR imaging with \JWST/NIRCam instrument. \citet{Harikane24} showed the \target is composed of multiple UV clumps with a total stellar mass estimated at $\sim 10^{9.6}$ \Msun. 

The paper is structured as follows. In \S~\ref{sec:obs} we present the \JWST/NIRSpec, NIRCam and ALMA observations, as well as data reduction of each of the datasets. In \S~\ref{sec:data_analysis} we describe the detailed analysis of each of the observations. In \S~\ref{sec:results}, we present and discuss our findings and in \S~\ref{sec:conclussions} we summarise the results of this work. Throughout this work, we adopt a flat $\Lambda$CDM cosmology: H$_0$: 67.4 km s$^{-1}$ Mpc$^{-1}$, $\Omega_\mathrm{m}$ = 0.315, and $\Omega_\Lambda$ = 0.685 \citep{2020A&A...641A...6P}.

\section{Observations and data reduction}
\label{sec:obs}

\subsection{NIRSpec Data}\label{sec:JWSTobs}

We observed \target with JWST/NIRSpec in IFS mode \citep{Jakobsen22,boker22}  as part of the GA-NIFS survey (PID 1217, PIs: S. Arribas \& R. Maiolino). The NIRSpec data were taken on 5\textsuperscript{th} of May 2023, with a medium cycling pattern of eight dither positions and a total integration time of 18.2~ks (5.05~h) with the high-resolution grating/filter pair G395H/F290LP, covering the wavelength range $2.87-5.27~\mu$m (spectral resolution $R\sim2000-3500$; \citealp{Jakobsen22}), and 3.9~ks (1.1~h) with PRISM/CLEAR ($\lambda=0.6-5.3~\mu$m, spectral resolution $R\sim30-300$).

Raw data files of these observations were downloaded from the Barbara A.~Mikulski Archive for Space Telescopes (MAST) and then processed with the {\it JWST} Science Calibration pipeline\footnote{\url{https://jwst-pipeline.readthedocs.io/en/stable/jwst/introduction.html}} version 1.11.1 under the Calibration Reference Data System (CRDS) context jwst\_1149.pmap. We made several modifications to the default reduction steps to increase data quality, which are described in detail by \citet{Perna23b} and which we briefly summarise here. Count-rate frames were corrected for $1/f$ noise through a polynomial fit. Furthermore, we removed regions affected by failed open MSA shutters during calibration in Stage 2. We also removed regions with strong cosmic ray residuals in several exposures. Any remaining outliers were flagged in individual exposures using an algorithm similar to {\sc lacosmic} \citep{vDokkum01}: we calculated the derivative of the count-rate maps along the dispersion direction, normalised it by the local flux (or by three times the rms noise, whichever was highest), and rejected the 95\textsuperscript{th} percentile of the resulting distribution \citep[see][for details]{DEugenio23ifs}. The final cubes were combined using the `drizzle' method. The main analysis in this paper is based on the combined cube with a pixel scale of $0.05''$.

\subsection{ALMA data}\label{sec:ALMAobs}

In this work, we use the ALMA programme 2018.1.00429.S, which contains higher-resolution observations of \CII emission line. The data were calibrated and reduced with the automated pipeline of the Common Astronomy Software Application (\texttt{casa}; \citealt{McMullin07}) version 5.6. We exclude two service blocks of the programme, as they were taken with significantly shorter baselines and hence giving a resolution of $\sim 0.5''$, giving a worse resolution in the combined reduced data

The data calibration of the ALMA data was performed using \texttt{CASA} v6.5.4. First, we subtract the continuum from the data using the \texttt{uvcontsub} task. We fit a linear model to the uv-visibilities using the channels without any emission-line contamination. We imaged the continuum-subtracted \CII visibilities and the dust continuum using \texttt{tclean} at 0.03 arcsec pixel scale using two separate weighting schemes. We first imaged the data using natural weighting to get the total flux from the \CII. This results in a final beam size of 0.45$\times0.5''$. We created a further data set using the Briggs weighting scheme with a robust parameter of 0.5. This resulted in a high-resolution data set with a beam size of 0.25$\times0.3''$. The dust-continuum maps and \CII emission line cubes were cleaned down to 3$\sigma$ with a 3$''$ circular mask on COS-3018. 

The \CII emission line cube was analysed using \texttt{Spectral cube} python library. We created zeroth, first and second-order \CII maps by collapsing the cube along the velocity range of -200 -- 200 \kms (i.e. $\pm$FWHM of the \CII emission line) centred on the systematic redshift of the \CII emission line.

\subsection{NIRCam imaging}\label{sec:imaging}

We use additional JWST/NIRCam imaging taken as a part of the PRIMER programme\footnote{\url{https://primer-jwst.github.io/}.} (PID 1837; PI J. Dunlop). These data
consist of imaging in eight NIRCam bands (F090W,
F115W, F150W, F200W, F277W, F356W, F410M and F444W) and we show the RGB image in top panel of Figure \ref{fig:prism_specs}.
The NIRCam images and photometry were obtained from the DAWN JWST Archive. These data were reduced using a combination of the \textsc{jwst} and \textsc{grizli}\footnote{10.5281/zenodo.1146904} pipelines \citep{Valentino23}.

\section{Data analysis}\label{sec:data_analysis}

\subsection{IFS data and emission line fitting}\label{sec:prep}

Before we can analyse the emission line cube of the PRISM or R2700 observations, we first need to perform a few preparatory steps: 1) masking of any outlier pixels not flagged by the pipeline; 2) background subtraction; 3) estimating the uncertainties on the data. For these tasks and the rest of the analysis, we use \texttt{QubeSpec}, an analysis pipeline written for NIRSpec/IFS data.

We need to mask any major pixel outliers that were not flagged by the data reduction pipeline. Although these pixels do not cause significant problems during the emission line fitting of galaxy-integrated spectra, these outliers can become a problem during spaxel-by-spaxel fitting. To identify the residual outliers not flagged by the pipeline, we used the error extension of the data cube. We flagged any pixels whose error is 10$\times$ above the median error value of the cube. We verified that this does not have any impact on the emission line maps by using the 5 and 20 thresholds without any changes to our conclusions. 

For both the PRISM and R2700 observations, we need to subtract the strong background affecting our observations. For the R2700 observations, we mask the location of the source based on its \OIIIL[5007] emission ($2\sigma$ SNR contours), and we estimate the background using \texttt{astropy.photutils.background.Background2D} (2D background estimator) task with 5$\times$5 spaxels box window, for each individual channel in the data cubes. We visually inspected the resulting background spectra and found no evidence of narrow features (e.g. emission or absorption lines). Therefore, to reduce noise, we smoothed the background in spectral space using a median filter with a width of 25 channels to reduce any noise effects. The final estimated background is subtracted from the flux data cube. 

The strong background in the PRISM observations requires subtraction of the background on the detector images and we employ the method described in \citet{Marconcini24}. Here, we briefly describe the procedure. The background in the PRISM observations is subtracted from the detector images for each of the dithers. Similarly to the R2700, we create a source mask based on the \OIIIL[5007] emission line with additional padding of 3 pixels. This mask is then deprojected to the 8 detector images, using the \textsc{blot} function in the \jwst pipeline. For each of the 2-d calibrated images, we fit the linear function to each of the slices in the dispersion direction, excluding the source pixels defined by our source mask. To filter out noisy features in the background, the estimated background is smoothed by a median filter with a width of 7 pixels. The final background-subtracted cube is constructed using the stage 3 pipeline step from the background-subtracted 2-D images.

\citet{Ubler23} reported that the uncertainties on the flux measurements in the \texttt{ERR} extension of the data cubes are underestimated, compared to the noise estimated from the rms of the spectrum, calculated inside spectral window free from emission lines. However, the error extension still carries information about the relative uncertainties between pixels and outliers. Therefore, when extracting each spectrum, whether it is a combined spectrum of multiple spaxels or directly fitting a spaxel, we first retrieve the uncertainty from the error extension. Then we scale this error extension uncertainty so that the error extension's median uncertainty matches the spectrum's sigma-clipped rms in emission line-free regions. This scaling is performed across both detectors independently, without a wavelength dependence.

\subsubsection{R2700 emission line fitting}\label{sec:R2700}

\begin{figure*}
        \centering
	\includegraphics[width=0.85\paperwidth]{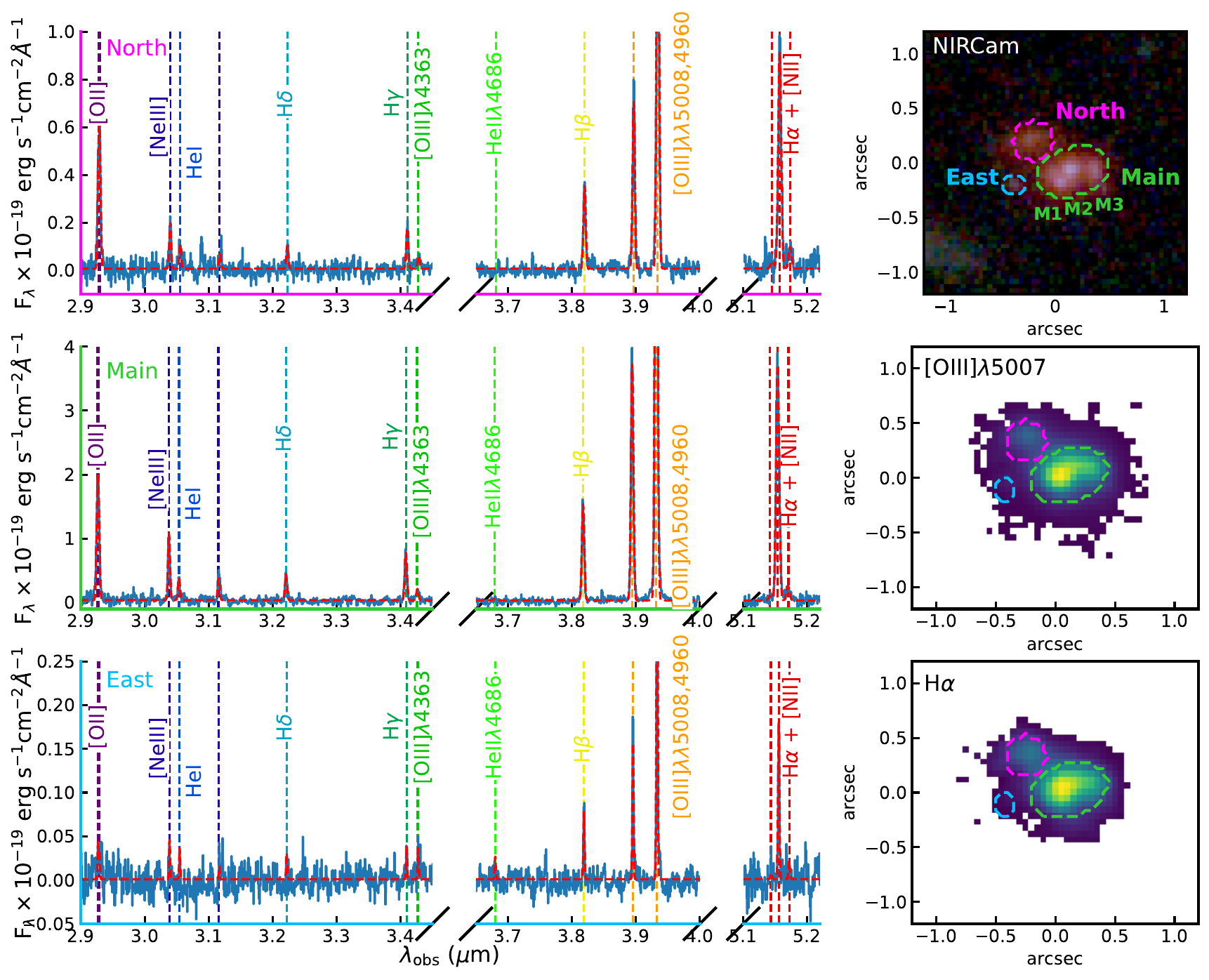}
    \caption{ Overview of the COS-3018 system. Right from top to bottom: RGB image from JWST/NIRCam imaging (with B - F115W, G-F200W, R-F444W filters), \OIIIL[5007] and \Halpha maps from R2700 NIRSpec/IFS cube. The coloured dashed line boxes indicate the regions used to extract the NIRSpec/R2700 spectra on the left. Left column: Full R2700 spectra extracted from the NIRSpec/IFU observations in black line.}
    \label{fig:R2700}
\end{figure*}

We initially fitted the integrated aperture R2700 spectra of the individual components (see Figure \ref{fig:R2700} and \S~\ref{sec:results}) as a series of Gaussian profiles for emission lines and power-law to describe the continuum, using the \texttt{Fitting} routines in \texttt{QubeSpec} code. In total we fit the following emission lines: \Halpha, H$\beta$, H$\gamma$, H$\delta$, \NIIall,\OIIIall, \OIIall, \NeIIIall, \HeIIL[4686] and HeI$\lambda$5875. We use a single Gaussian per emission line as our main model. We tie the redshift (centroid) and intrinsic FWHM of each Gaussian profile to a common value to reduce the number of free parameters, leaving the flux of each Gaussian profile free. For each emission line, the FWHM of the line is convolved with the line spread function of NIRSpec from the JDOCS
\footnote{Available at \href{https://jwst-docs.stsci.edu/jwst-near-infrared-spectrograph/nirspec-instrumentation/nirspec-dispersers-and-filters}{jwst-docs website}.}.

We fixed the \OIIIL[5007]/\OIII[4959] flux ratio to be 2.99 \citep{Dimitrijevic07}; and \NIIL[6583]/\NIIL[6548] to 3.06 (based on the atomic transition probability; \citealt{Osterbrock06}). Lastly the [O~{\sc ii}]$\lambda$3727/[O~{\sc ii}]$\lambda$3729 \space to vary between 0.69 and 2.6 to reflect the dependence of such doublet on the electron density \citep{Sanders16}.

Furthermore, each spectrum is also fitted with a 2-Gaussian model, which includes an additional Gaussian profile in \Halpha, H$\beta$ and \OIIIall emission lines. We only fit an additional "narrow-2" component in these strong emission lines, because they are the only lines with sufficient signal-to-noise ratio to detect non-Gaussian line profiles arising from complex kinematics or outflows. We will further discuss this additional component in \S~\ref{sec:kinematics} \& \ref{sec:outflows}. We use the BIC
\footnote{The Bayesian Information Criterion \citep{Schwarz78}, which uses $\Delta \chi^{2}$ but also takes into account the number of free parameters, by penalising the fit for more free parameters. BIC is defined as BIC=$\Delta \chi^{2} + k \log(N)$, where N is the number of data points and k is the number of free parameters.} 
to choose whether the fit needs a second narrow component (using $\Delta$BIC $>10$ as a boundary for choosing a more complex model).

The fiducial model parameters for the single and multi-component models are estimated with a Bayesian approach, where the posterior probability distribution is calculated using the Markov-Chain Monte-Carlo (MCMC) ensemble sampler - \texttt{emcee} \citep{emcee, foreman-mackey+2013}. For each of the variables, we need to define set priors for the MCMC integration. The prior on the redshift of each spectrum is set as a truncated Gaussian distribution, centred on the systemic redshift of the galaxy with a sigma of 300 km s$^{-1}$ and boundaries of $\pm 1000$ km s$^{-1}$. The prior on the intrinsic FWHM of the narrow-line component is set as a uniform distribution between 100-500 km s$^{-1}$, while the prior on the amplitude of the line is set as a uniform distribution in logspace between 0.5$\times$rms of the spectrum and the maximum of the flux density in the spectrum. For the second Gaussian component in the strong emission lines, the velocity offset is set as a truncated Gaussian distribution with mode 0 and with a sigma of 250 km s$^{-1}$ and boundaries of $\pm$1000 km s$^{-1}$, while for the FWHM of the outflow component, we use a uniform distribution with boundaries of 500-2000 km s$^{-1}$. 

The final best-fit parameters and their uncertainties are calculated as median value and 68 \% confidence interval of the posterior distribution. We note that all the quantities derived from R2700 spectral fitting (e.g. gas densities, temperatures, outflow velocities and metallicities) are calculated from the posterior distribution to account for any correlated uncertainties in the spectrum.

\begin{figure*}
        \centering
	\includegraphics[width=0.85\paperwidth]{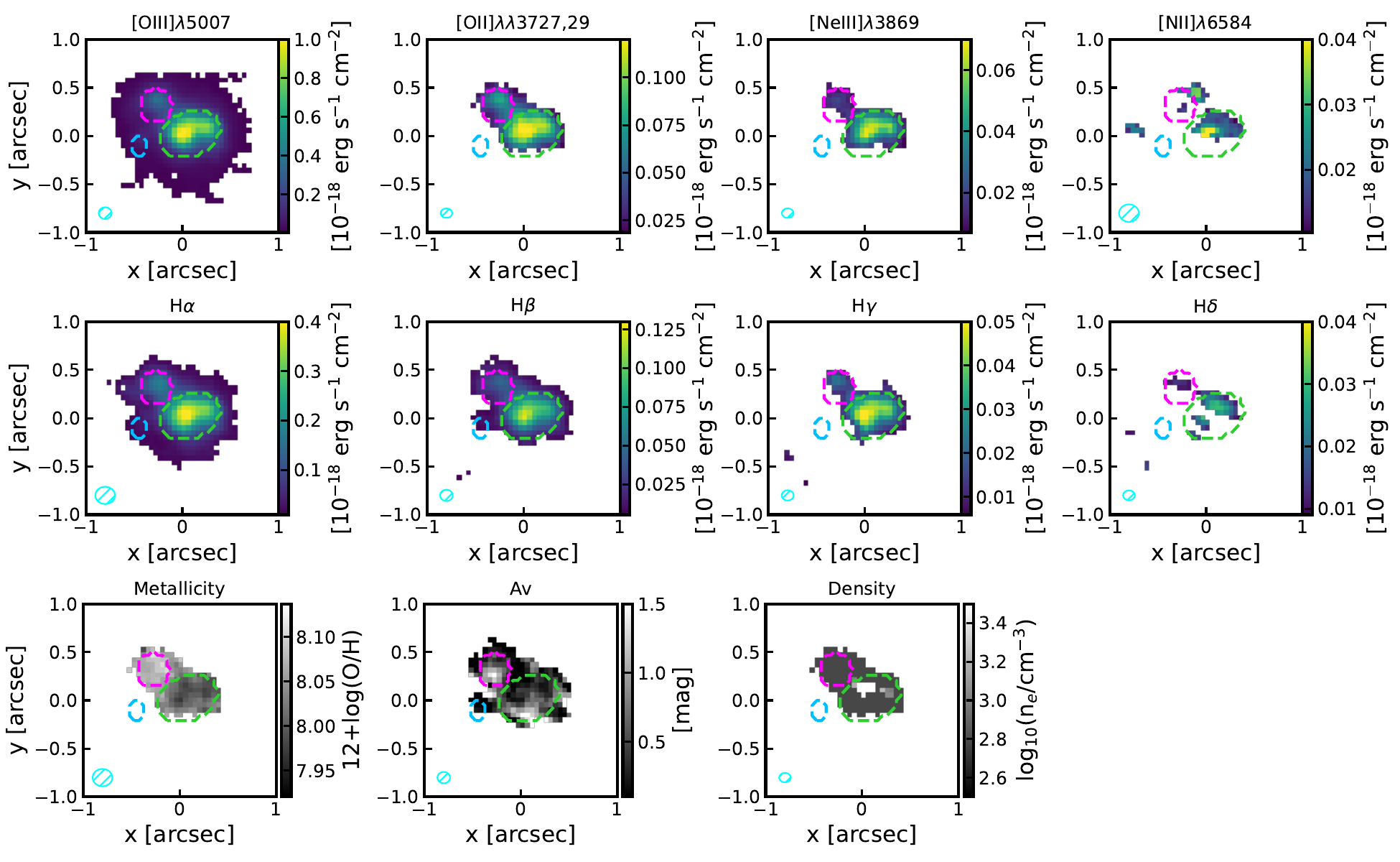}
    \caption{Resolved emission line maps of \target and derived ISM physical properties from the R2700 NIRSpec IFs cube. In each map, we also show the Main, North and East components as green, purple and light blue contours, respectively. Top row: Forbidden lines detected in \target from left to right:\OIIIall, \OIIall, \NeIIIall and \NII emission lines. Middle row: detected Balmer lines in the target: \Halpha, H$\beta$, H$\gamma$ and H$\delta$. We also detect HeI line in the combined spectra, however, we do not detect it in the individual spaxel spectra. Bottom row: Spatially resolved ISM properties from the strong emission lines, from left to right: Metallicity from strong line calibrations, dust attenuation (from \Halpha \& \Hbeta) and electron density. The cyan hatched region indicates the size of JWST/NIRSpec PSF at the wavelength of the emission line.}
    \label{fig:Flux_maps}
\end{figure*}

\begin{figure}
        \centering
	\includegraphics[width=0.9\columnwidth]{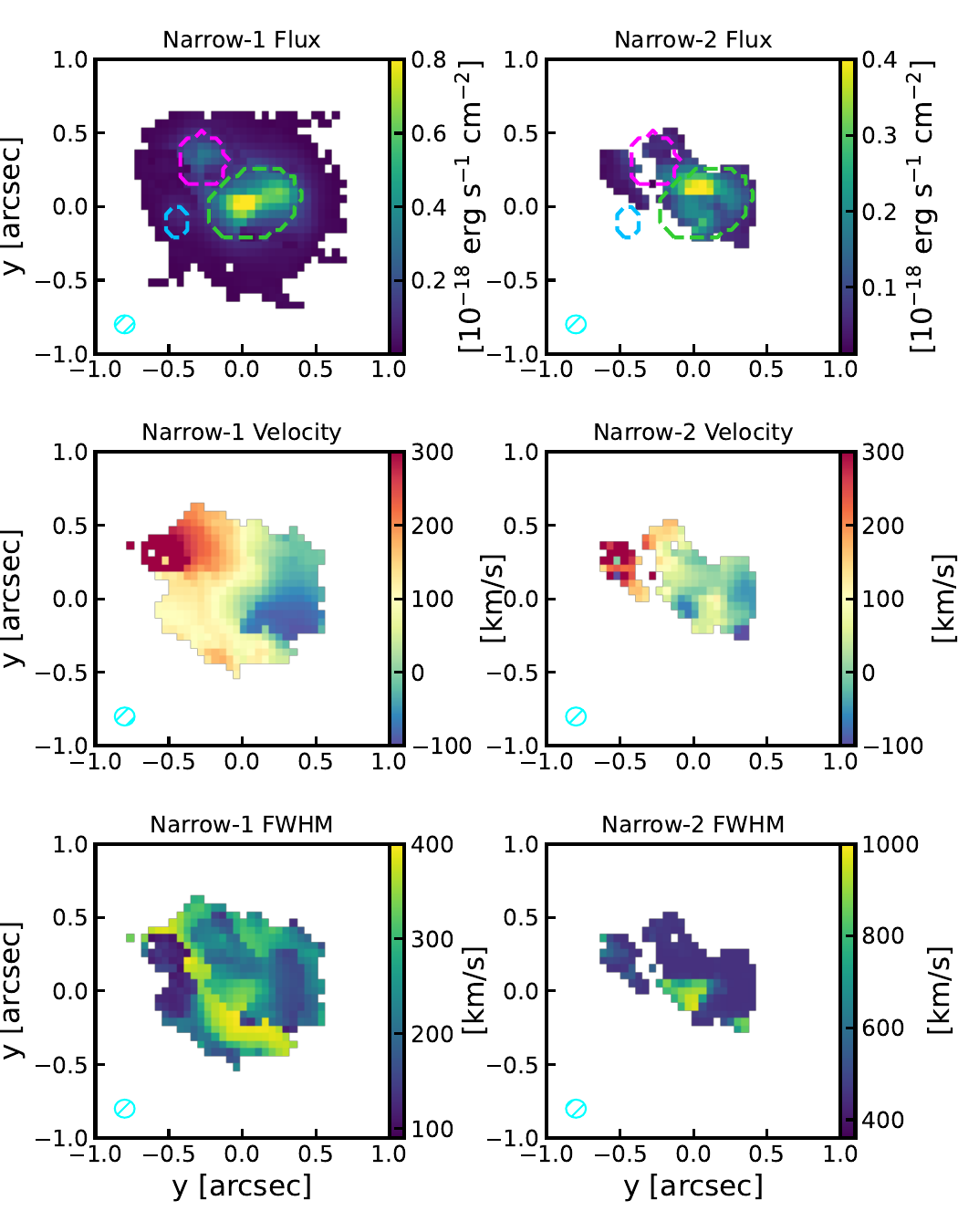}
    \caption{Kinematics of the \OIIIL[5007] emission for the narrow-1 and narrow-2 components shown in left and right columns, respectively. From top to bottom: Flux, velocity and FWHM for each of the components. Velocity maps are caluclated relative to the redshift of the Main component. In each map, we also show the Main, North and East components as green, purple and light blue contours, respectively. The cyan hatched ellipse indicate the \JWST/NIRSpec PSF at the wavelength of the \OIIIL[5007]. \OIIIL[5007] and the \Halpha have the same kinematics when fitted independently.}
    \label{fig:vel_maps}
\end{figure}

We repeat the spectral fitting on spaxel-by-spaxel in the region covered by the target, using the same models as for the integrated spectra described above. We do not do any PSF matching or spaxel binning at this point. As outlined above, the \OIIIall, \Halpha and \Hbeta have a complex emission line profile which requires multiple Gaussian profiles. In order to describe the emission line profile we use a non-parametric description of the emission line profile: v10, v50 v90, and W80 parameters described as velocity containing 10, 50, 90 \% of the flux and velocity width containing 80 \% of the flux (v90-v10), respectively. We will use the W80 parameter to describe velocity dispersion of the emission line profile \footnote{For a Gaussian profile  W80 = 1.033$\times$ FWHM = 2.427$\times \sigma$.}. We show the final derived spatially resolved flux maps in Figure \ref{fig:Flux_maps} and kinematics in Figures \ref{fig:vel_maps} \& \ref{fig:kinematics}. We note that the kinematics map are the same if we fit all lines simultaneously or we fit \OIIIL[5007] and \Halpha separately. 

\begin{figure*}
        \centering
	\includegraphics[width=0.85\paperwidth]{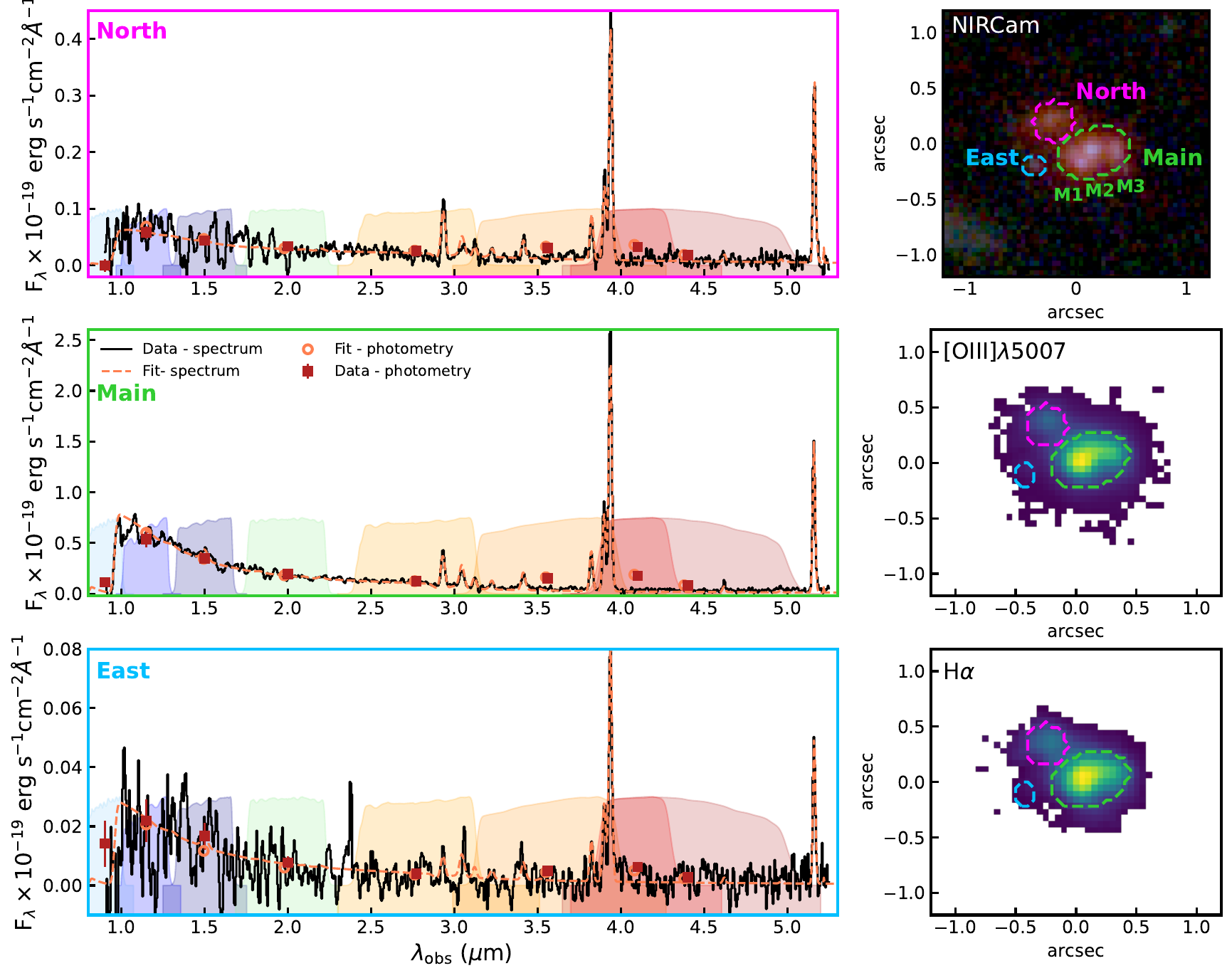}
    \caption{ Overview of the COS-3018 system. Right from top to bottom: RGB image from JWST/NIRCam imaging (with B - F115W, G-F200W, R-F444W filters), \OIIIL[5007] and \Halpha maps from R2700 NIRSpec/IFS cube. The coloured dashed line contours indicate the regions used to extract the NIRSpec/PRISM spectra on the left. Left column: Full NIRSpec/PRISM spectra extracted from the NIRSpec/IFU observations in black line. We overlay the NIRCam photometry (from the dashed line apertures) as red points with the NIRCam transmission curve as various coloured shaded regions.}
    \label{fig:prism_specs}
\end{figure*}

The uncertainties on the measured quantities are derived by estimating the input from the posterior distribution from the \texttt{QubeSpec} fitting code to derive the physical quantities. The final value and uncertainties on the properties are estimated as the median value and standard deviation. However, we note that the majority of the derived quantities are dominated by the systematic uncertainties from the individual calibrations used in this work. 

\subsubsection{PRISM fluxes}\label{sec:PRISM}

To fit the emission lines in NIRSpec/PRISM (see Figure \ref{fig:prism_specs}), we employed \texttt{pPXF} \citep{Cappellari2017, Cappellari2022}, to fit the complete stellar continuum and emission lines simultaneously. The full description of this procedure is reported in \citet{DEugenio24}. The continuum is fitted as a linear superposition of simple stellar-population (SSP) spectra, using non-negative weights and matching the spectral resolution of the NIRSpec/PRISM observations \citep{jakobsen_nirspec_2022}. For the stellar templates, we used the synthetic library of simple stellar population spectra (SSP) from \textsc{fsps} \citep{conroy+2009,conroy_gunn_2010}. This library uses MIST isochrones \citep{choi+2016} and C3K model atmospheres \citep{conroy+2019}. We also used a 5\textsuperscript{th}-order multiplicative Legendre polynomial, to capture the combined effects of dust reddening, residual flux calibration issues, and any systematic mismatch between the data and the input stellar templates. To simplify the fitting, any flux with a wavelength shorter than the Lyman break is manually set to 0. 

For the emission lines fitting, we use the redshift determined from the NIRSpec/R2700 observations as an initial value. All emission lines are modelled as single Gaussian functions, matching the observed spectral resolution. In order to remove degeneracies in the fitting and reduce the number of free parameters, the emission lines are split into two separate kinematic groups, bound to the same redshift and \emph{intrinsic} broadening. These groups are as follows:
\begin{itemize}
    \item UV lines with rest-frame $\lambda< 3000$~\AA. 
    \item optical lines with rest-frame $3000 <\lambda< 7000$~\AA. 
\end{itemize}

As described in \S~\ref{sec:R2700}, we fixed the emission line ratio to the value prescribed by atomic physics (e.g., \OIIIL[5007]/\OIIIL[4959]= 2.99). For multiplets arising from different levels, the emission line ratio can vary. In addition, as the \HeIIL[1640] and OIII]$\lambda\lambda$1661,66 are blended, we fit them as a single Gaussian component. We report the measured fluxes for each of the components in Table \ref{table:eml}.

\subsection{SED modelling}\label{sec:SED_fit}

For the SED fitting, we simultaneously fit the PRISM spectroscopic and NIRCam photometry, as the UV continuum of the fainter components is not detected in the spectroscopic data but it is detected in the NIRCam imaging. We use \texttt{Prospector} v2.0 \citep{Johnson21}. Before we fit the data, we PSF matched the NIRCam data and PRISM IFU observations to the PSF of NIRSpec/IFU at \Halpha wavelength. \texttt{Prospector} is a Bayesian SED modelling framework built around the stellar-population synthesis tool \textsc{fsps} \citep{Conroy2009, Conroy2010}. To set up the model, we used a non-parametric star-formation history (SFH), consisting of constant SFR in pre-defined time bins. We employ a `continuity' prior between the individual SFH bins (this prior penalises sharp changes in SFR between adjacent time bins; see \citealt{Leja2019} for more details). In total, we use eight SFH bins with the two most recent bins being 10 and 50~Myr, which is then followed by 6 equally spaced bins in log space between 1000~Myr and $z=20$ (no stars are formed earlier). The prior on the stellar mass and metallicity has a flat log distribution, while dust attenuation is described by a flexible dust attenuation law, consisting of a modified Calzetti law \citep[][]{Calzetti00} with a variable power-law index \citep{Noll2009} tied to the UV-bump strength \citep{Kriek2013}. Stars younger than 10~Myr are further attenuated by an extra dust screen, parametrised as a simple power law \citep{charlot00}. Overall, the parameters of the host galaxy follow the setup of \citet{Tacchella22}, including coupling ongoing star formation to nebular emission using pre-computed emission-line tables \citep{Byler17}.

As we are fitting a number of data sets simultaneously, we need to use the noise `jitter' term (on the spectrum only), which can scale the input noise vector by a uniform factor (with flat prior between 0.5 and 2). The PRISM spectrum has multiple noisy features in the blue and red ends of the spectrum. As such we mask the upturn in the spectrum reward of 5.25 $\mu$m and blueward 0.8 $\mu$m. The posterior distribution of our model parameters is estimated using nested sampling \citep{Skilling2004}, implemented using the \textsc{dynesty} library; \citep{Speagle2020,Koposov23}. 

The final results of the emission line fitting are summarised in Table \ref{table:SED} and we show the best fits the spectroscopy and photometry in Figure \ref{fig:prism_specs} and Figures \ref{fig:prosp_main}, \ref{fig:prosp_north} and \ref{fig:prosp_east}. 

\section{Results \& Discussion}\label{sec:results}

In this section, we present and discuss our results based on the analysis outlined above. In \S~\ref{sec:sys} we describe this complex system, in \S~\ref{sec:agn} we search for any presence of an AGN, and we present results of the SED fitting in \S~\ref{sec:sed_results}. We investigate the ISM properties and oxygen and nitrogen abundances in \S~\ref{sec:ism}. In \S~\ref{sec:kinematics} \& \ref{sec:outflows}, we investigate the kinematics and presence of outflows in \target. Finally in \S~\ref{sec:ALMA} we make a comparison with the \jwst and ALMA \CII observations.

\subsection{Description of the system}\label{sec:sys}

The new NIRCam and NIRSpec/IFS data showed that this galaxy is a complex system, comprising at least three components: Main, North and East. We show these components as green, magenta and light blue contours in NIRCam RGB, H$\alpha$ and \OIIIL[5007] images in Figures  \ref{fig:prism_specs} and \ref{fig:R2700}. Furthermore, the Main component has multiple separate clumps clearly seen at the shorter wavelengths (< 2$\mu$m) that we dub M1, M2 and M3 and we show these in NIRCam image in Figures  \ref{fig:prism_specs} and \ref{fig:R2700}. The three UV peaks remain barely resolved at longer wavelengths (such as \Halpha or F444W filter) due to the lower instrumental spatial resolution at longer wavelengths. Indeed, in \OIIIL[5007], the M1 and M2 are blended. To investigate the properties of the three components, we extracted the NIRCam photometry and NIRSpec/PRISM and R2700 spectra from the same regions as defined by coloured regions in the right panels of Figure \ref{fig:prism_specs}. We show the extracted NIRCam photometry and  NIRSpec/PRSIM spectra in Figure \ref{fig:prism_specs} and NIRSpec/R2700 spectra in Figure \ref{fig:R2700}.

All three components are detected in \Halpha, \Hbeta, \OIIIall, \NeIIIall and H$\gamma$, additionally, we detect \NIIall and \OIIIL[4363] in the Main and North component. The \NIIall detection is interesting in particular as this is currently one of the highest redshift detections of this emission line. For example in the JADES survey \citep{DEugenio24}, this emission line is very rarely detected above z$>4$, despite some observations reaching almost 40h integration time with the NIRSpec/MSA using the efficient R1000 grating. The velocity offset between the Main and North components is 302$\pm$5 \kms, while the velocity offset between the Main and East components is 120$\pm$3 \kms. We will discuss whether these components belong to the same galaxy or whether this is a major merger in \S~\ref{sec:kinematics}. 

We also define two \NIIL[6584] regions based on the \NIIL[6584] emission line maps (see top right panel of Figure \ref{fig:Flux_maps}) and we dub them as Main-\NII and North\NII regions and we show the extracted spectra for each of the regions in Figure \ref{fig:NII_clumps}. We will further discuss these regions in \S~\ref{sec:agn} \& \ref{sec:ism}.

This galaxy was initially selected based on \textit{HST}+\textit{Spitzer} photometry based on its high EW of \OIIIall + \Hbeta of 1424$\pm$143 $\AA$. We measured the equivalent width of these lines from the PRISM spectroscopy for the sum of all components as well as individual lines. We find the EW for the whole system of 1620$\pm$160 $\AA$, agreeing within 1$\sigma$ with the photometric results. These extremely high EWs result in \target being selected as an Extreme Emission Line Galaxy (EELG) by \citet{Boyett24}.

We measured a UV spectral slope ($\beta_{UV}$) of the Main and North components of -2.02$\pm0.05$ and -1.31$\pm0.21$, respectively. Unfortunately, due to poor SNR of the continuum in the PRISM observation of the East component, we are not able to measure a reliable $\beta_{UV}$ from spectroscopy. Comparing these measurements to the median $\beta_{UV}$ value from the NIRCam observations in the JADES survey of  $\beta_{UV}=-2.26\pm0.03$ \citep[][]{Topping24}, the Main component is similar with other galaxies at similar redshifts (e.g., \citealt{Dunlop13,Bowler14}). The North component has a very high (red) $\beta_{UV}$, indicating large dust attenuation in the North component as has been reported in a few cases at z$\sim$7 \citep{Smit18}. We will further investigate the dust content in these components in \S~\ref{sec:ALMA}. 

\begin{table*}
   \caption{Emission line fluxes of the three main components identified in \S~\ref{sec:sys} from both the PRISM and R2700 grating data. The fluxes are in the units of $10^{-19}$ \ergscm. The upper limits are set as $3\sigma$. For \Halpha and \NIIall and \OIIall we only report the combined flux for the PRISM as we are unable to deblend the emission lines.}
   \centering
 \input{Table_fluxes}
  \label{table:eml}
\end{table*}

\subsection{Searching for AGN}\label{sec:agn}
The extreme \OIIIL[5007] EWs and excess of \OIIIL[88$\mu$m] compared to \CII led previous studies to speculate that COS-3018 could host an AGN \citep{Witstok22}. Therefore, in this section, we investigate any evidence of AGN in this complex system based on the new \jwst spectroscopy of rest-frame UV and optical emission lines. 

The simplest approach to search for AGN is through any emission from the broad line region (BLR), dense ionised clouds orbiting close to the supermassive black hole, resulting in broad emission (FWHM$>10^3$\,\kms) components in permitted lines, typically Balmer hydrogen lines (\Halpha and H$\beta$;  e.g., \citealt{Maiolino23JADES}) without any counterpart in the strongest forbidden lines (such as \NIIL[6584], \OIIall and \OIIIall). We extracted the spectra of each of the sub-systems and fitted them with a multi-component model. Unlike for the fiducial fits (see \S~\ref{sec:data_analysis}), we do not tie the kinematics of the broad components of the permitted lines with the forbidden lines. We do not see any evidence of a broad component in the Balmer lines with different kinematics compared to the forbidden lines. This shows that any broad component in permitted lines is also seen in the forbidden lines, tracing a medium with sub-critical density, associated with outflows (see \S~\ref{sec:outflows}) rather than a BLR. Furthermore, we extracted a spectrum from the three UV clumps in the Main system (i.e. M1, M2 and M3) and we do not see any evidence of any component in the 'forbidden' lines that are not seen in the 'permitted' lines (see Figure \ref{fig:outflow}). Based on this analysis we do not see any evidence for a type-1 AGN in this complex system, even an offset AGN such as is the case in the study by \citet{Ubler24b}. 

\begin{figure*}       
	\includegraphics[width=0.8\paperwidth]{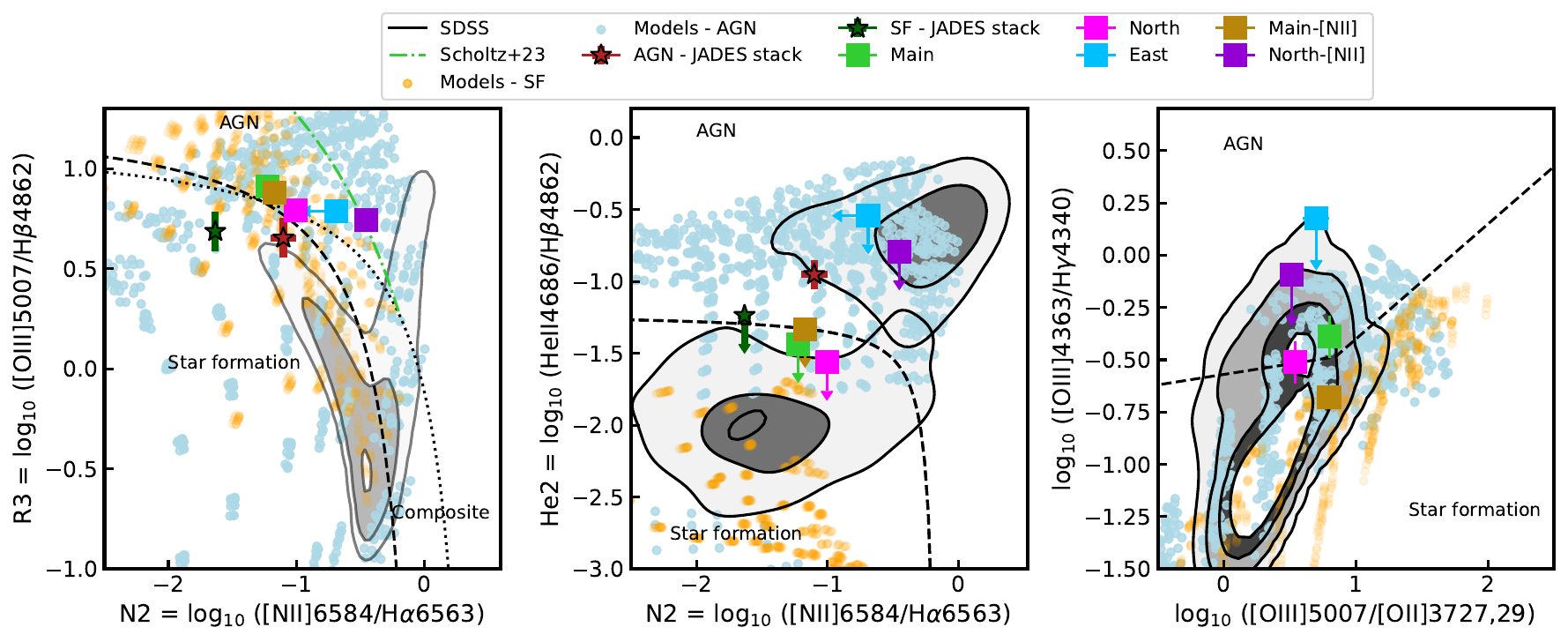}
    \caption{ Optical emission line diagnostics sensitive to the source of ionisation in COS-3018. Left: N2-R3 BPT (\NII/\Halpha \space vs \OIII/\Hbeta; top row). We show the demarcation lines between star formation and AGN from \citet{Kewley01}, \citet{Kauffmann03} and \citet{Scholtz23AGN} as black dashed, dotted and green dash dotted lines, respectively.  Middle: He2-N2 (\HeIIL[4686]/\Hbeta \space vs \NII/\Halpha) diagram The black dashed line indicates a demarcation line between star-forming and AGN galaxies by \citet{Shirazi12}. The black contours show the star-forming galaxies and AGN from SDSS, respectively. Right panel: \OIIIL[4363]/H$\gamma$ vs \OIIIL[5007]\OIIall. The black dashed line indicates the demarcation line from \citet{Mazzolari24}. In each panel, we show ionisation models from \citet{Feltre16,Gutkin16} as yellow and light blue points, respectively. The magenta and cyan squares show a stacked spectrum for AGN and star-forming galaxies from 
    \citet{Scholtz23AGN}. }
    \label{fig:AGN_optical}
\end{figure*}

In order to determine the presence of a type-2 AGN, we investigate any presence of high ionisation lines such as \NeIV, \NeV or \HeIIL[1640] or \HeIIL[4686] as well as using common emission line ratio diagnostics such as BPT. As we do not detect the \NeIV nor \NeV emission line, we investigate 
this system using the \OIII/\Hbeta vs \NII/\Halpha (BPT; \citealt{BPT1981}), \HeIIL[4686]/\Hbeta vs \NII/\Halpha (He2-N2; \citealt{Shirazi12, Tozzi23}) and new diagnostics using \OIIIL[4363] from \citet{Mazzolari24} in Figure \ref{fig:AGN_optical}. In each diagram, we plot the Main, North and East components as green, magenta and cyan squares, respectively. Furthermore, we also show the photo-ionisation models from \citet{Feltre16,Gutkin16} for AGN and star-forming galaxies and SDSS galaxies as contours. 

As described in e.g. \citet{Scholtz23}, the standard BPT diagram is no longer able to distinguish between AGN and star-forming galaxies in lower mass and low metallicity galaxies such as those at high redshift. Indeed, all components except for North-\NII region lie in AGN region of the BPT diagram based on the original demarcation lines from \citet{Kewley01} and \citet{Kauffmann03}; however, they are consistent with star-formation based on the new demarcation line from \citet{Scholtz23} which  rules out objects with low \NIIL/\Halpha and high R3 as low metallicity star-forming galaxies with high specific star-formation rates (sSFR; SFR/stellar mass). The smaller North-\NII region is on the demarcation line and cannot be reliably established as being consistent with only AGN ionisation.  

In the middle panel of Figure \ref{fig:AGN_optical}, we investigated the \HeIIL[4686]/\Hbeta vs \NIIL[6584]/\Halpha, an alternative diagnostic proposed by \citet{Shirazi12} and recently applied e.g.\, by \citet{Tozzi23, Scholtz23, Ubler23}. We do not detect \HeIIL[4686] in any of the three components in COS-3018. The upper limits on the \HeIIL[4686] in the Main and North components lie in the star-forming part of this diagram, while the upper limit for the East component lies in the AGN diagram. 

Finally, we investigated the new diagnostic diagram from \citet{Mazzolari24}, involving the \OIIIL[4363]. We plot \OIIIL[4363]/H$\gamma$ vs \OIIIL[5007]/\OIIall in the right panel of Figure \ref{fig:AGN_optical}. The Main and North components have a strong (over 5$\sigma$) detection of \OIIIL[4363] and they lie on the line separating AGN and star-forming galaxies from \citet{Mazzolari24}. We do not see any major differences in the line ratios, suggesting that the detection of \OIIIL[4363] is driven by the high surface brightness of \OIII in that region, rather than elevated \OIIIL[4363] emission line ratios.

Overall, the ionisation properties of \target are consistent with both AGN and star-formation and we do not see any definitive evidence that there is a type-2 or type-1 AGN in either of the three components of \target. We stress, however, that at the Epoch of Reionisation, signatures of AGN can be easily hidden by the young stellar population \citep{Tacchella24}. Therefore, the peculiar ionisation conditions described in \citet{Witstok22} can be driven purely by star formation rather than AGN activity. For the rest of this work, we will treat the emission of this galaxy as originating from a starburst rather than the narrow line region of an AGN.

\subsection{SED fitting results}\label{sec:sed_results}

To derive basic stellar population properties, we utilised SED fitting using \texttt{prospector} to model simultaneously the NIRCam photometry and NIRSpec/PRISM spectra. We extracted the photometry and PRISM spectra from regions defined in \S~\ref{sec:sys} and we show the extracted data and the best fit to both spectra and photometry in Figure \ref{fig:prism_specs}. We show the full fits along with the posterior distributions and star-formation histories in the Appendix in Figures \ref{fig:prosp_main}, \ref{fig:prosp_north} and \ref{fig:prosp_east}. The final values for the fitted parameters are summarised in Table \ref{table:SED}.

The SED fitting showed that the Main and North components are currently undergoing a starburst, with an increased SFR in the past 10 million years by a factor of more than 5-10 compared to the previous 100 Myr. The Main component is the most massive with log$_{10}$(M$_{*}$/\Msun)=9.4$\pm$0.1, the North component has log$_{10}$(M$_{*}$/\Msun)=9.2$\pm$0.1 and the East component is the least massive one with log$_{10}$(M$_{*}$/\Msun)=7.7$\pm$0.2. The total stellar mass of the system islog$_{10}$(M$_{*}$/\Msun)=9.7$\pm$0.1. Each of the components has a low estimated gas-phase metallicity of $\sim$20\% Z$_{\odot}$. 
The estimated mass of these objects is higher than those estimated from the \textit{HST}+ \textit{Spitzer} by $\sim$0.5 dex \citep{Bouwens15} and within 1$\sigma$ of the NIRCam photometry only \citep{Harikane24}. 

\begin{figure*}
        \centering
	\includegraphics[width=0.8\paperwidth]{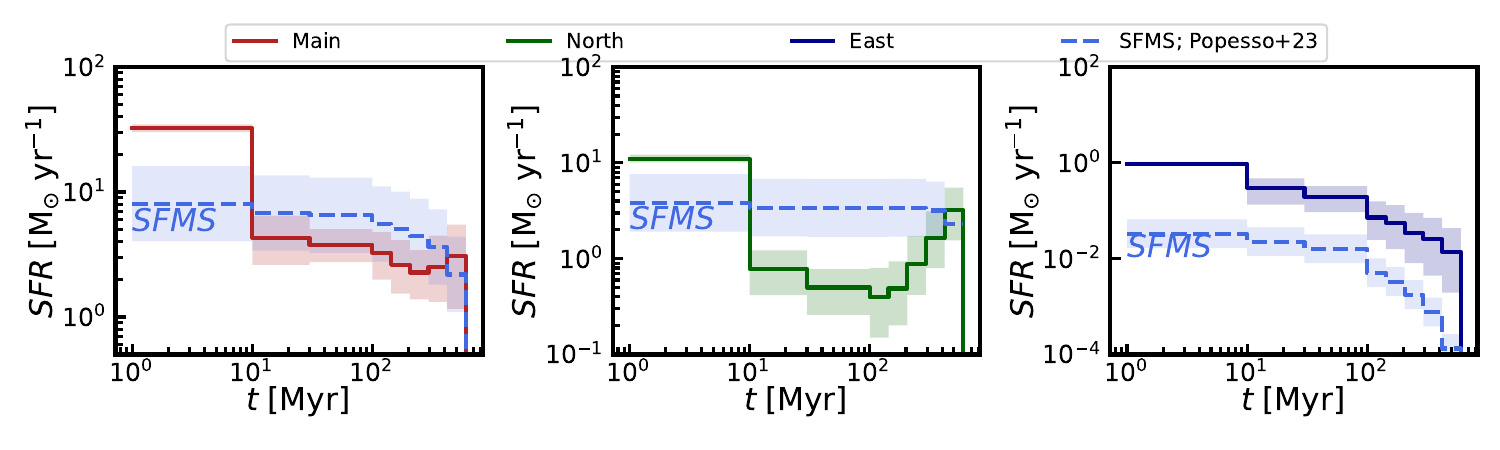}
    \caption{ Star formation histories of the three clumps in COS-3018. From left to right: Main, North and East components respectively. The time in the x-axis is defined from start of the SFH, with t= 0 Myr corresponds to the redshift of the source. We plot the star-forming main sequence from \citet{Popesso23} as a blue dashed line for comparison. The Main and North components are going through a major starburst, being $\sim$10 higher than the main sequence.}
    \label{fig:sfh}
\end{figure*}

The SFRs averaged over the past 10 million years are 31, 11 and 0.5 \Msun yr$^{-1}$ for the Main, North and East components, respectively. This is significantly less than those derived from dust corrected \Halpha flux of 95, 23 and 2 \Msun yr$^{-1}$, respectively. There are numerous reasons for the discrepancy in SFRs from \Halpha and SED fitting. \Halpha emission probes SFR on timescales of <$5$Myr while the smallest SFR bin in our SED fitting is 10 Myr. Furthermore, the calibrations used to estimate SFR from \Halpha \citep{Kennicutt12} assume a constant SFR for the past 100 Myr, clearly not applicable to \target (see Figure \ref{fig:sfh}). Lastly, the \Halpha flux (and \OIII for that matter) can be boosted by the presence of a type-2 AGN, which is not ruled out based on our AGN diagnostics (see \ref{sec:agn}). 

The deep NIRSpec/PRISM and NIRCam data allow us to probe the star-formation history of this system. We plot the derived SFH in Figure \ref{fig:sfh} for each of the three main components along with the star-forming main sequence from \citet{Popesso23} as blue dashed lines. All three components are currently going through a star-bursting phase, starting about 10 Myr ago. While the Main component has been forming stars for the past 400 Myr, the star-formation histories indicate that the North component has been forming stars for $\sim$300 Myr, followed by a brief 100 Myr break until 10 Myr ago.

\begin{table}
   \caption{Results of the SED fitting of the NIRCam photometry and PRISM spectroscopy along with the results of the R2700 NIRSpec spectroscopy.}
   \centering
     \renewcommand{\arraystretch}{1.3}
 \begin{tabular}{lccc}
  \hline
  \hline
  Component & Main & North & East \\
  \hline
  Prospector SED fitting \\
  \hline
  M$_{\rm UV}$ &  -20.80$^{+0.04}_{-0.04}$ &-19.30$^{+0.04}_{-0.04}$ & -18.11$^{+0.04}_{-0.04}$\\
  log$_{10}$(Mass/M$_{\odot}$) & 9.35$^{+0.09}_{-0.12}$ & 9.17$^{+0.07}_{-0.07}$& 7.67$^{+0.16}_{-0.19}$\\
  log$_{10}$(SFR$_{10}$/M$_{\odot}$yr$^{-1}$) & 1.51$^{+0.03}_{-0.03}$ & 1.07$^{+0.05}_{-0.06}$ & -0.04$^{+0.04}_{-0.04}$ \\
  log$_{10}$(SFR$_{100}$/M$_{\odot}$yr$^{-1}$) & 0.84$^{+0.06}_{-0.07}$ & 0.07$^{+0.05}_{-0.06}$ & -0.70$^{+0.10}_{-0.10}$ \\
  12+log(O/H) & 8.01$^{+0.03}_{-0.03}$ & 8.0$^{+0.03}_{-0.02}$ & 8.0$^{+0.03}_{-0.02}$\\
  log$_{10}$(U) & -1.72$^{+0.02}_{-0.01}$ &-2.21$^{+0.05}_{-0.04}$ &-1.48$^{+0.16}_{-0.16}$ \\
  A$_{\rm v}$ (SED) & 0.23$^{+0.05}_{-0.04}$ &0.71$^{+0.06}_{-0.07}$ &0.12$^{+0.05}_{-0.05}$ \\
  \hline
  Emission lines - R2700\\
  \hline
  log$_{10}$(SFR/M$_{\odot}$yr$^{-1}$); \Halpha$_{\rm corr}$ & 1.98 & 1.36 & 0.36 \\
  12+log(O/H) (strong )(N2) & 8.05$^{+0.01}_{-0.01}$ & 8.15$^{+0.01}_{-0.01}$& 7.96$^{+0.05}_{-0.05}$ \\
  12+log(O/H) (strong) (no N2) & 8.00$^{+0.01}_{-0.01}$ & 7.88$^{+0.11}_{-0.11}$& 7.93$^{+0.06}_{-0.06}$ \\
  12+log(O/H)(\Te) & 8.17$^{+0.07}_{-0.08}$ & 7.9$^{+0.1}_{-0.1}$ &- \\
  log$_{10}$(N/O) &-1.20$^{+0.05}_{-0.04}$ & -1.16$^{+0.06}_{-0.07}$& <-0.98\\
  \Te \OIIIL[4363](K)& 12 750$\pm$720 &16 000$\pm$1400 & - \\
  n$_{e}$ (cm$^{-3}$) &1217$\pm$250 & 711$\pm$210& - \\
  A$_{\rm v}$ (\Halpha/\Hbeta) & 0.43$^{+0.06}_{-0.06}$ & 0.62$^{+0.09}_{-0.07}$ & 0.00$^{+0.42}_{-0.00}$\\
  \hline
 \end{tabular}
  \label{table:SED}
\end{table}

\subsection{Interstellar medium properties}\label{sec:ism}

The forest of emission lines detected in COS-3018 allows us to perform a detailed analysis of the ISM in the three different components. Using both nebular and auroral lines in the R2700 spectrum we can perform a detailed study of chemical abundance patterns in this galaxy, employing the `direct', Te-method. 


We detect \OIIIL[4363] in the Main and North components with SNR of 5-9, which allows us to constrain the electron temperature of the \OIII emitting gas. We employ \texttt{python's} \textsc{pyneb} library for chemical abundances, using the atomic data from the \textsc{chianti} database. The full description of the procedure is in \citet{Curti24_9.4}. 

To estimate the ISM properties, we use the total narrow line emission line profile, that we attribute to the galaxy emission (see \S~\ref{sec:kinematics}). We corrected the line fluxes for dust attenuation. We estimated the dust attenuation (A$_{\rm v, gas}$) using the \Halpha and \Hbeta assuming SMC extinction law and Case B recombination, which is appropriate for high-z galaxies \citep{Curti24_9.4}. We also mapped the A$_{\rm v}$ (see bottom row of Figure \ref{fig:Flux_maps}) and we will further discuss this map and its comparison to the ALMA dust emission map in \S~\ref{sec:ALMA}.

We derive the temperature of the \OIII emitting gas (O$^{++}$) exploiting the high SNR detection of \OIIIL[5007] and \OIIIL[4363] in the R2700 spectra. We simultaneously also derive the gas density using the \OII doublet ratio whose lines are very well spectrally resolved in the R2700 data. The temperature of O$^{+}$ emitting region (hereafter \Tii) is 
assumed in the process to follow the temperature-temperature relation from \cite{izotov_low_2006}, i.e. $t_{2} = 0.693\xspace t_{3} + 2810$, where the $t_{2}$ and $t_{3}$ are the temperatures of the O$^{+}$ and O$^{++}$ emitting gas, respectively. 

For the Main component, we infer a temperature of the O$^{++}$ emitting gas of $t_{3}$~=~$12800\pm 730$ K, while the gas density is $n_{e}=1240\pm240$ cm$^{-3}$, which is consistent with typical density derived for high-redshift galaxies \citep[$\sim$500--1200 cm$^{-3}$; e.g.][]{isobe_density_jwst_2023, Marconcini24, Lamperti24,Rodriguezdelpino24}. For the North component, we derived a significantly lower $n_{e}$ of $710\pm210$ cm$^{-3}$ with temperature of the O$^{++}$ emitting gas of $t_{3}$~=~$16140\pm 1430$ K. In the East component, we do not detect \OIIIL[4363] and hence we are unable to derive the  O$^{++}$ temperature as done for the other components, however, we are able to derive a $n_{e}$ from the \OII of $1200\pm1100$ cm$^{-3}$, assuming a temperature of 15 000 K.

We took advantage of the spatially and spectrally resolved \OIIall doublet and estimated the density across the system (see Figure \ref{fig:Flux_maps}). We detect \OIIall in spaxel-by-spaxel analysis in the Main and North components. Although we detect \OIIall in the East component in the integrated region spectrum, we do not spatially resolve it. The estimated density map peaks in the central UV clump in the Main component with a density of 9450$\pm$2000 cm$^{-3}$. We verified this feature in the spatially resolved $n_{e}$ map by extracting an integrated spectrum and verifying the estimated value. 

With the derived temperature and density of the O$^{++}$ gas, we can now estimate the relative ionic abundances of oxygen and hydrogen using the intensity $I(\lambda)$ for each of the species, while taking into account the different temperature- and density-dependent volumetric emissivity of the transitions $J$:
\begin{equation}
    \frac{N(X^l)}{N(Y^m)} = \frac{I_{(\lambda)l}}{I_{(\lambda)m}} \frac{J_{(\lambda)m} (T,n)}{J_{(\lambda)l} (T,n)} \ .
\end{equation}

\begin{figure}
        \centering
	\includegraphics[width=0.99\columnwidth]{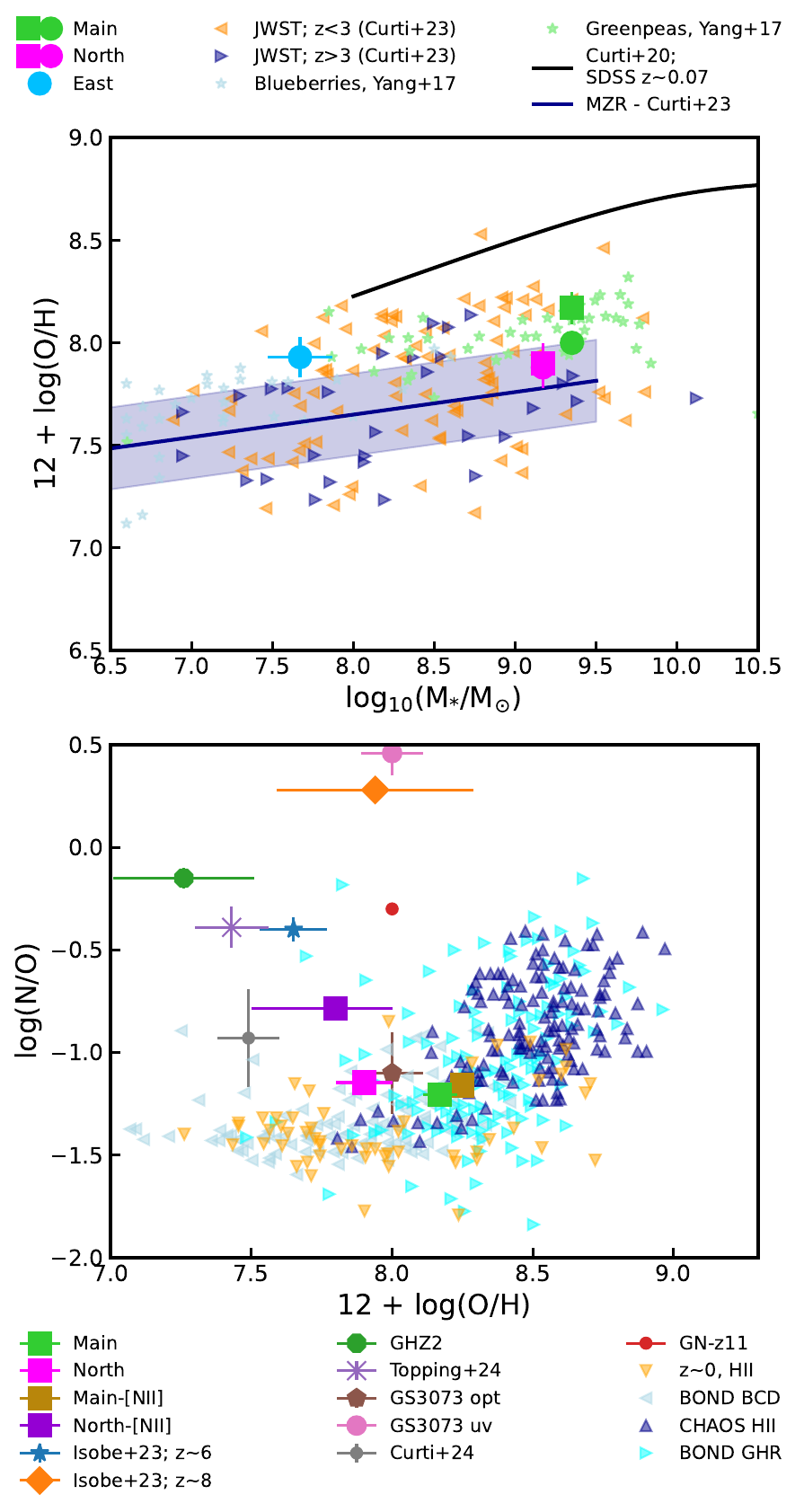}
    \caption{Top: location of \target on mass-metallicity relation (MZR). The green, magenta and blue points show Main, North and East components for metallicity measurement using T$_{e}$ (squares) and strong line calibration (circles) methods. The orange and dark blue triangles show \jwst metallicity measurements of \jwst galaxies from \citet{Curti23}. The blue solid line and the shaded region indicate the mass-metallicity relation at z$>$3 \citep{Curti23}. The black solid line shows the mass-metallicity relation from SDSS at z=0.07 \citep{Curti20}. The light blue and green stars show Blueberries and Greenpeas from \citet{Yang17a, Yang17b}.
    Bottom: Comparison of oxygen and nitrogen abundance of different components of \target. The red, blue, gold and purple squares show the Main, North components, Main-[NII] and North-[NII] clumps, respectively. The circles show data from \jwst \citep[][]{Cameron23gnz11,Isobe23,  Castellano24, Curti24_9.4,Ji24,Topping24}. The triangles show comparison of local data HII regions \citep{Tsamis_HII_2003, esteban_reappraisal_2004, esteban_keck_2009, esteban_carbon_2014, esteban_abundance_2017, garcia_rojas_2004, garcia_rojas_2005, garcia_rojas_2007, peimbert_2005, garcia_rojas_esteban_2007, lopez_sanchez_2007, toribio_san_cipriano_2016, toribio_san_cipriano_carbon_2017}; local dwarfs \citep{berg_carbon_2016, berg_chemical_2019, vale_asari_bond_2016, senchyna_2017}.}
    \label{fig:metallicity}
\end{figure}

Using the outlined method, we derive the O$^{++}$/H and O$^{+}$/H from the \OIIIall/\Hbeta (assuming t=\Tiii) and \OII/\Hbeta (assuming t=\Tii), respectively, and compute the total oxygen abundance as O/H = O$^{+}$/H + O$^{++}$/H.
In our calculation, we do not take into account the abundance of O$^{3+}$. This species has an ionisation potential of $54.9$eV, which is extremely close to the ionisation potential of HeII. Given no detection of HeII, we can assume that its contribution to oxygen abundance is negligible. 
We derived a 12+log(O/H) for Main and North components of $8.17\pm0.07$ and $7.9\pm0.1$, respectively. We compared \target to the rest of the galaxy population on the mass-metallicity plane (MZR-plane) as green and magenta squares in the top panel of Figure \ref{fig:metallicity} with other \JWST studied galaxy and local galaxies from SDSS as well as local analogues of low metallicity high-z galaxies (blueberries and green peas, \citealt{Yang17a, Yang17b}). 

As we do not detect \OIIIL[4363] on a spatially resolved basis or in the East component, we use the strong line calibration from \citet{Curti20} to explore spatial metallicity variation across the galaxy. As we are mainly interested in relative differences between the individual components, our analysis is robust against the systematic differences in the absolute metallicity values as introduced by the choice of the specific calibration set.
We calculate the gas-phase metallicities based on the following emission line ratios: O3O2=\OIIIL[5007]/\OIIall, R23= (\OIIIall+ \OIIall)/\Hbeta and \NeIIIL/\OIIall. We use N2= \NIIL/\Halpha ratio for the integrated spectra from individual regions, however, we do not use it for the spatially resolved metallicity maps as it is rarely detected in individual spaxels.

For the integrated spectra from the individual components, we estimate a 12+log(O/H) of 8.05$^{+0.01}_{-0.01}$, 8.15$^{+0.01}_{-0.01}$ and 7.96$^{+0.05}_{-0.05}$ for Main, North and East components, respectively, when we include N2 and 8.00$^{+0.01}_{-0.01}$, 7.88$^{+0.11}_{-0.11}$ and 7.93$^{+0.06}_{-0.06}$ without N2 emission line ratio. We plot the O/H abundances derived from strong line calibrations in the top panel of Figure \ref{fig:metallicity} for the Main, North and East components as green, magenta and cyan circles, respectively. We summarised these results in Table \ref{table:SED}.

The North component of \target is consistent with the MZR derived for galaxies at $z>3$ from \citet{Curti23} (using the same set of strong-line calibrations) and with the Greenpeas from \citet{Yang17a} (derived with the \Te-method). However, the Main and the East components are $\sim$0.4--0.5 dex higher than the MZR showing signs of significant metal enrichment in these components at z$\sim$6.85. 

We repeated the analysis on the spaxel-by-spaxel basis, only using spaxels where we detect \OIIall and \NeIIIall emission lines without including any information from \NIIL[6584] and we show the metallicity map in the bottom left panel of Figure \ref{fig:Flux_maps}. Overall, we see an increasing metallicity gradient towards the North component, consistent with the values derived for the individual components without \NII.

\subsubsection{N/O abundance}\label{sec:N_O}

The detection of \NIIall in \target is currently one the highest redshift detections of this emission line \citep[see][for a detection at $z\sim6.9$]{Arribas23} and we can constrain the nitrogen abundance of this galaxy. We detect \NIIall in the integrated spectra of the Main and North components (see top and middle panels of Figure \ref{fig:R2700}) as well as in the two locations in the \NIIL[6584] maps (see top right panel of Figure \ref{fig:Flux_maps}). Interestingly, we do not detect either NIII]$\lambda$1750 nor NIV]$\lambda$1492 emission lines used in the literature to derive N/O abundance \citep{Cameron23,Isobe23, Curti24_9.4, Ji24, Schaerer24, Topping24}, despite the strong detection of \NIIL[6585] in \target. As described above, we do not see any evidence of strong hard ionising radiation that would ionise nitrogen to higher species such as N$^{++}$ or N$^{+++}$, which require 29.6 and 47.45 eV, respectively. Furthermore, the sensitivity of our observations in the PRISM spectroscopy is only $<3\times 10^{-18}$ erg/s/cm2, too high to detect these lines.

As we do not detect NIII]$\lambda$1750 nor NIV]$\lambda$1492 emission lines in the NIRSpec/PRISM observations, we rely on the \NIIall detection to constrain N/O abundance. We use the calibration from \citet{Hayden-Pawson22} using the \NIIL[6585]/\OIIall emission line ratio, specifically their equation 1. We measured a log(N/O) abundance of -1.21$\pm$0.04 and -1.17$\pm$0.06 for the Main and North components. We show the derived N/O in the bottom panel of Figure \ref{fig:metallicity} along with a comparison to local HII regions, galaxies at z=0-2 as well as the latest results from \jwst spectroscopy at high-z \citep{Cameron23gnz11, Isobe23, Topping24, Curti24_9.4, Ji24}. All the components and regions of \target lie on the relation between O/H and N/O expected from low redshift galaxies and HII regions. 

Interestingly, we also detect \NIIL[6585] in the spatially resolved maps (see top right panel of Figure \ref{fig:Flux_maps}). The brightest  \NIIL[6585] clump is located on the brightest \OIII peak (from now on called Main-\NII clump), while another \NIIL[6585] clump is located in the 0.3 arcseconds West of the peak of the North component (North-\NII clump). We verified the reliability of spatially resolved maps by extracting integrated spectra encompassing the \NII clumps (see Figure \ref{fig:NII_clumps}), recovering the total flux from the maps. There is no difference between the Main-\NII clump and the Main component in the\NIIL[6585]/\Halpha (see left panel of Figure \ref{fig:AGN_optical}). On the other hand, the North-\NII clump has an elevated \NIIL[6585]/\Halpha ratio by 0.4 dex compared to both the Main and North components. 

The \OIIIL[4363] is only marginally detected in the North-\NII clump ($\sim$3.34$\sigma$) and we measure a temperature of 11000$\pm$1200 K and 12+log(O/H) (using the T$_{e}$ method) of 7.6$^{-0.2}_{+0.2}$, while the strong line calibration, excluding N2 from \citet{Curti20} estimates 8.06$^{-0.03}_{+0.03}$. We use an average of the strong line calibration value and the direct T$_{e}$ method with an error describing the full range of the values and uncertainties. Plotting the two \NIIL[6585] clumps on the N/O vs O/H plane (bottom panel of Figure \ref{fig:metallicity}), we see elevated N/O of the North-[NII] clump compared to Main and North components.

A PRISM spectrum extracted from the Main-\NII and North-\NII regions does not show any detection of UV emission lines such as NIII]$\lambda$1750 nor NIV]$\lambda$1492, \CIIIall, \CIVall and \HeIIL[1640]  emission lines, similar to the rest of \target. However, the increased N/O in the small faint region suggests that the origin of the increased N/O seen in high-z galaxies can originate from individual small off-centred regions in the galaxy. This is intriguing given the recent finding of high N/O abundances at low metallicities discovered in some high-z galaxies by JWST. These have been often associated with the formation of proto-globular clusters in the early Universe \citep{Dantona2023,Cameron23gnz11,Belokurov_kravtsov_nitrogen_2023,Ji2024_nitrogen}. Therefore, this small (consistent with a point source), nitrogen-rich and metal poor offset clump may be tracing a globular cluster in the process of formation in the halo of the main galaxy.

\subsection{Kinematics}\label{sec:kinematics}

In this section, we will explore the ionised gas kinematics of \target. \citet{Smit18} have identified this object as a rotating disk based on ALMA observations of \CII emission (resolution of $\sim0.4$ arcsec) and it was further characterised as a turbulent disk \citep[see][]{Parlanti23}. The JWST/NIRCam imaging and JWST/NIRSpec spectroscopy showed multiple extended components in this system with masses of $>10^{9}$ \Msun, too large to be individual star-forming clumps. Furthermore, the individual clumps have different star-formation histories as well as ISM properties such as ionisation parameter, density and metallicity, suggesting a different origin of the components. This would imply a merger of multiple galaxies, rather than a single rotating galaxy.

\begin{figure*}
        \centering
	\includegraphics[width=0.8\paperwidth]{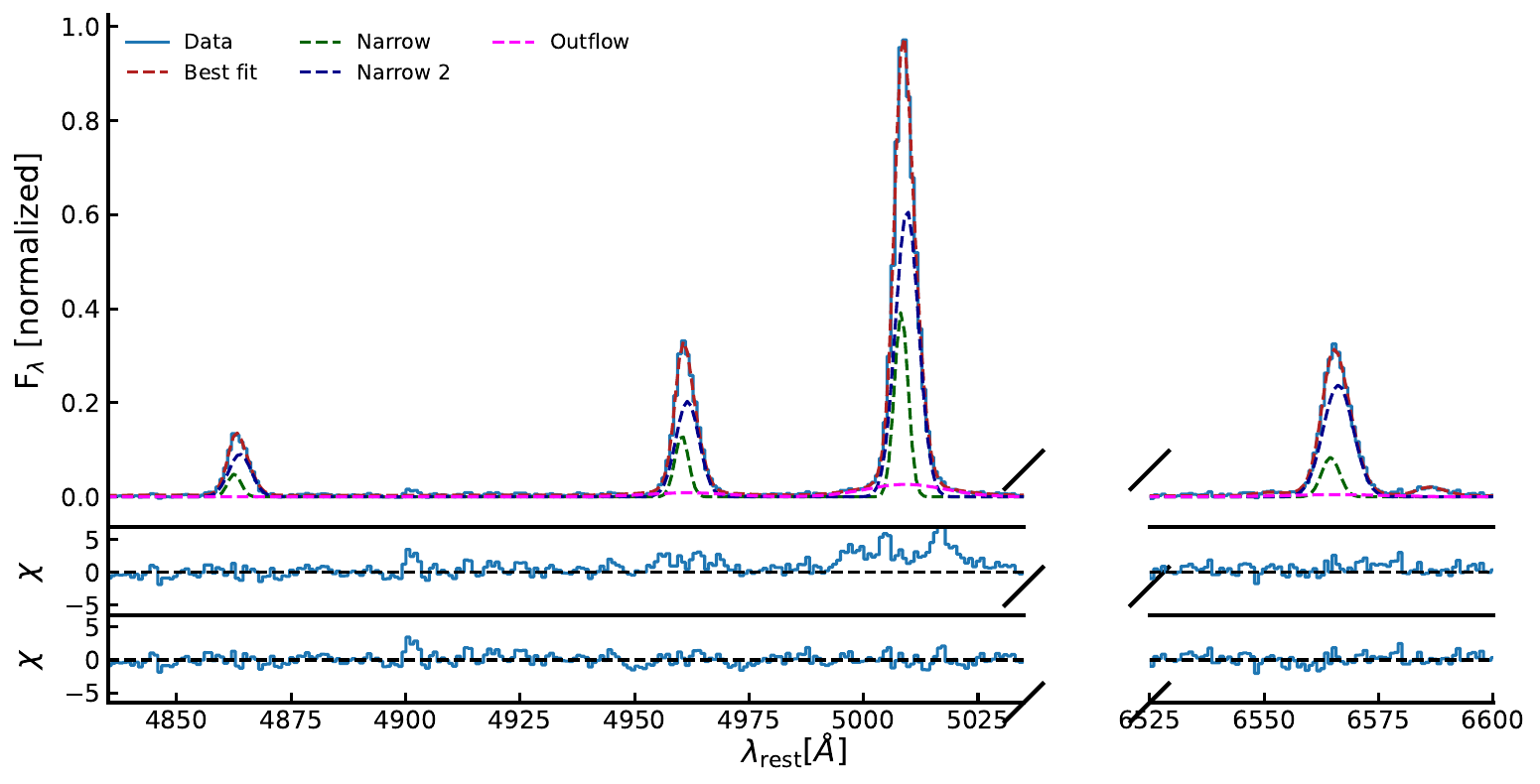}
    \caption{Modelling the emission line profile of the [OIII]5007, H$\beta$ and \Halpha to identify any broad outflow components in the Main component. Top panel: The spectrum of the Main component and the best-fit model. The data is shown as a light blue line. The green, blue and magenta dashed lines show the two narrow components in [OIII], Balmer lines (\Halpha and H$\beta$) and the outflow components respectively. The red dashed line shows the total model. Middle panel: Residuals for a model without the broad outflow component. Bottom panel: Best fit residuals with the full model including a broad outflow component. }
    \label{fig:outflow}
\end{figure*}

The extracted spectra from the individual \target components can only be reproduced by fitting multiple Gaussian profiles in both \Halpha, H$\beta$ and \OIIIall emission lines. We show the full decomposition of the emission line profile in the Main component in Figure \ref{fig:outflow}. In total, we fit three Gaussian to the \OIIIall, \Halpha and H$\beta$, while only a single Gaussian component to \NIIall, due to its limited detection. In the Main component, the "narrow-1" and the "narrow-2" components (green and dashed lines in Figure \ref{fig:outflow}) have FWHM of 210$\pm$12 and 350$\pm$6 km s$^{-1}$, respectively. We spatially map both the narrow and narrow-2 components and we find that the velocity gradient of the "narrow-2" component is aligned with the kinematics of the narrow component (see Figure \ref{fig:vel_maps}). We also model a third broad Gaussian component in the 1D spectra of each of the spectra, which we dub the "outflow" component and we will discuss its origin in \S~\ref{sec:outflows}. 

\begin{figure*}
        \centering
	\includegraphics[width=0.9\paperwidth]{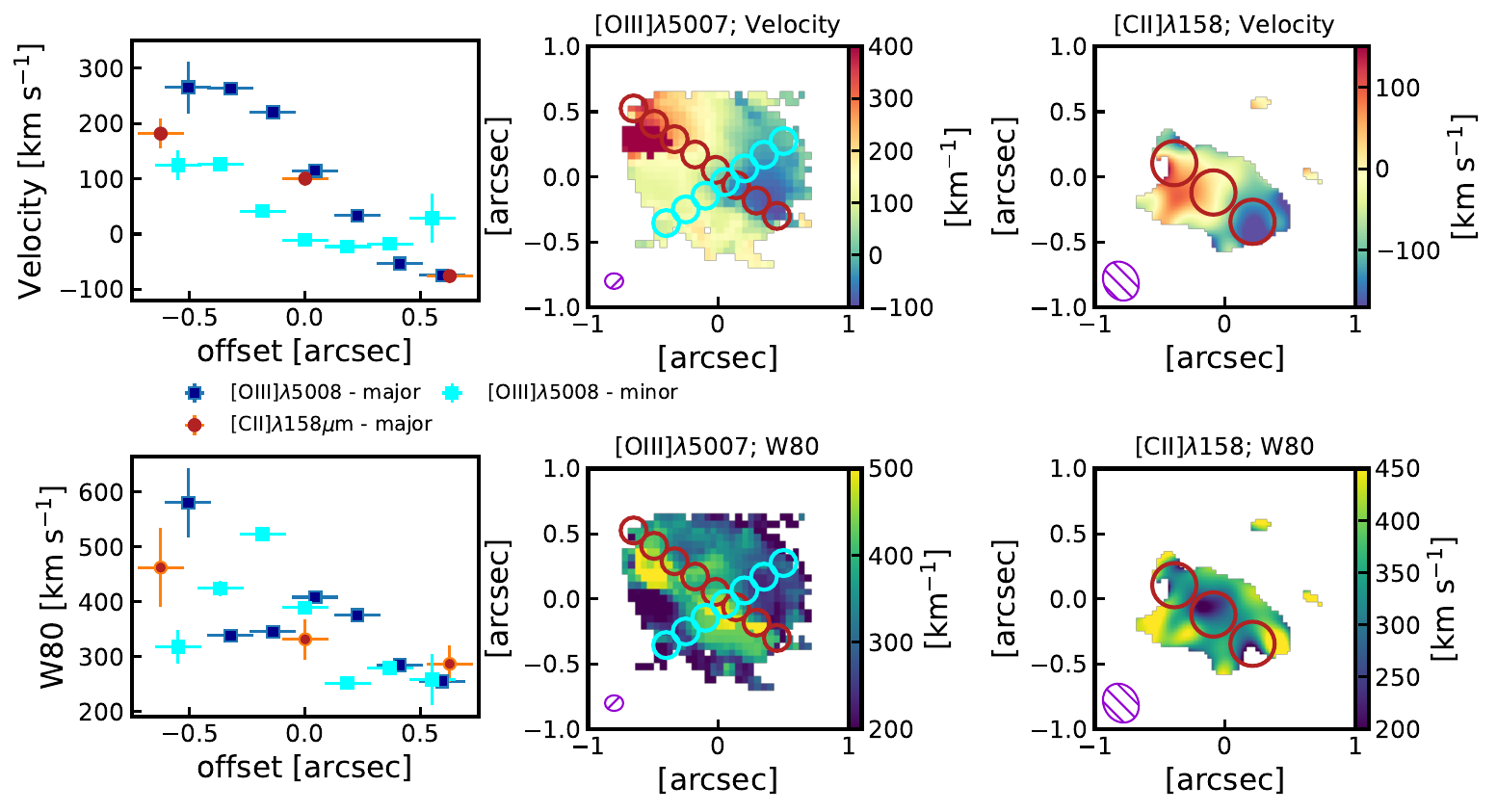}
    \caption{ Results of our kinematical analysis of \target for both \OIIIL[5007] and \CII. Top row from left to right: Position-velocity diagram, velocity maps for \OIIIL[5007] and \CII (bottom panel). Same as top row but for velocity width (W80; velocity width containing 80\% of the emission line flux). In each of panel, magenta  hatched circles show the PSF of the JWST/NIRSpec and ALMA observations. The apertures used to extract the position-velocity data are indicated as red and cyan for major and minor axis. }
    \label{fig:kinematics}
\end{figure*}

Overall, it is difficult to disentangle the two components ("narrow" and "narrow-2"; see Figure \ref{fig:vel_maps}) and hence we use the v50 and W80 parameters (see \S~\ref{sec:R2700} for definition) to describe the overall emission line profile. The bulk of the emission line profile is tracing the kinematics of the ionised gas from the galaxy components, however, given the high values of the W80, it is likely that some low-velocity outflows contaminate this overall profile as well (see Figure \ref{fig:kinematics}).  

We show the v50 velocity and W80 maps (corrected for instrumental broadening and excluding the "outflow" component) for \OIIIall and \CII in the middle and right columns of Figure \ref{fig:kinematics}. In both \OIIIall and \CII, we see the largest velocity gradient in the northeast-southwest direction between the Main and North components defined in \S~\ref{sec:sys}. The W80 map shows high velocity width in the \OIIIall and \CII gas with W80 of up to 600 \kms. Furthermore, the peak velocity map shows velocity gradients in multiple directions: between the Main and North components and the Main and East components. 

In order to compare the kinematics traced by the \OIIIL[5007] and \CII, we extracted spectra along a pseudo slit centred on the brightest \OIIIL[5007] peak and aligned with the largest velocity gradient. We show the individual apertures used to extract the spectra as red circles in the peak velocity and W80 maps of \OIIIL[5007] and \CII. 

We show the extracted position-velocity (PV) diagram along the NE-SW direction in the left column of Figure \ref{fig:kinematics} for \OIIIL[5007] and \CII emission lines as blue and dark orange points. We centered the pseudo-slit on the brightest \OIIIL[5007] clump. The extracted peak velocity and W80 agree across the \OIIIL[5007] and \CII emission lines. However, the factor $\sim$2 improvement in PSF in JWST/NIRSpec observations compared to the ALMA observations as well as improved sensitivity allows us to map the extended ionised gas in this galaxy. Overall the system has multiple components (Main, North and East) with high velocity offsets of over 400 \kms and high velocity widths (W80) of 300-500 \kms, unlikely tracing a rotating disk. Indeed, observations and simulations of galaxies at z$\sim$2-7 showed that merging galaxies can have a smooth velocity gradient as seen in \target \citep[e.g.][]{Simons19, Rizzo22, Jones24}.

Furthermore, extracting velocity information at $\sim$80 degrees to the largest velocity gradient (i.e. tracing the second largest gradient along its minor axes, see the purple apertures in figure \ref{fig:kinematics}) also reveals a smaller velocity gradient. The velocity width of the line peaks off the centre of the Main component. This can be potentially a rotation in the individual UV clumps (M1,M2 and M3; see Figure \ref{fig:prism_specs}) in the Main component extended towards the East component (likely a low mass companion).

Additional information to distinguish between a major merger and a rotating disk is the measured stellar masses of the individual components. Both the Main and North components have stellar masses of $10^{9.1}$ \Msun making them too massive to be individual smaller star-forming clumps in \citep[e.g. ][]{Zanella18}.

Our findings suggest that this is a merger of two high-mass galaxies (Main and North components) with a low-mass companion (East). We utilise a definition of 'close pair' of galaxies from \citet{duncan19,Ventou19,Romano21,Mestric22,Perna23b}, which requires the galaxies to be within $\Delta r \lesssim20$\,kpc and $\Delta$v$\lesssim500$\,\kms. Given that the velocity offset of the North and East components is of 303 and 120 \kms, and the size of the entire \target is $\sim$7 kpc across, all three components meet this definition and these will most likely merge in the future. 
Similar scenarios are seen in other z$>$3 systems observed within the GA-NIFS survey \citep{Jones24, Lamperti24, Rodriguezdelpino24}.  

Given this analysis, it is difficult to interpret this object as a rotating disc, and it is most likely a merger of at least three different components. The merger interpretation of the kinematics also sheds light on young star-bursting SFH we observed in this system (see \S~\ref{sec:SED_fit}). The large molecular gas reservoir of 1.2$\pm0.2\times 10^{10}$\Msun estimated from \CII emission line \citep{Witstok22} shows that this is a gas-rich merger triggering star-burst episodes in the Main and North components. 

\subsection{Outflows}\label{sec:outflows}

In this section, we investigate the presence of any outflows in \target and their effect on the galaxy. As described above, the overall emission line profile of the Main component requires three Gaussian profiles.  The third Gaussian component (dubbed as "outflow", magenta line in Figure \ref{fig:outflow}) is significantly broader than the other ones, with FWHM of 1210$\pm$120 \kms. These high velocities are unlikely to come from the merger kinematics and it is more likely tracing an outflow. The outflow component has a velocity offset of only 55$\pm$26 \kms with respect to the narrow Gaussian components, i.e. well centred on the emission from the galaxy. We discuss what is driving this outflow later in this section. It is important to discuss the origin of the "narrow-1" and "narrow-2" components. These are possibly due to non-rotating motion due to the ongoing merger, as well as being contaminated by outflows. However, as it is impossible to decompose from the complex kinematic of the merger, we will only concentrate on the broadest component in this section. 

\begin{figure} 
        \centering
	\includegraphics[width=0.9\columnwidth]{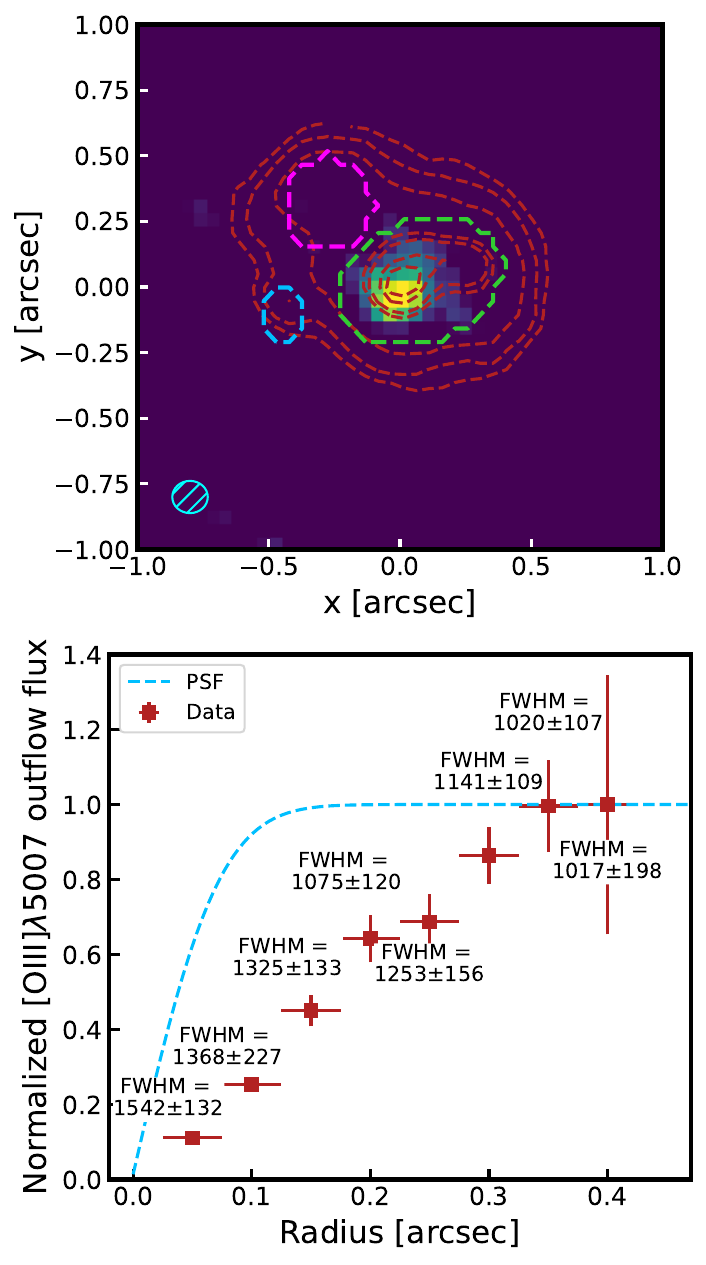}
    \caption{Top panel: Map of the \OIII  outflow with red contours showing the unperturbed  \OIII. We collapse the NIRSpec/R2700 cube around the spectral region of the outflow, excluding the channels affected by the galaxy emission. The green, magenta and blue dashed contours show the Main, North and East components. The cyan hatched circle shows the NIRSpec/IFU PSF at \OIIIL[5007] wavelength from \citet{DEugenio23ifs}. Bottom panel: Curve-of-growth for the flux of the broad outflow component normalized to the total flux. The data points are plotted as red squares, while the blue dashed line shows the NIRSpec/IFU PSF at the wavelength of \OIIIL[5007]. At each point, we also show the FWHM of the outflow component.}
    \label{fig:outflow_spatial}
\end{figure}

To find the location and extent of the outflow, we first map the outflow. We collapsed the continuum subtracted NIRSpec R2700 IFU cube over the channels centred on the velocity range $\pm1200$ \kms\ around \OIIIL[5007] excluding channels including the narrow \OIIIL[5007] emission.

We show the outflow map in the top panel of Figure \ref{fig:outflow_spatial} with red contours showing the narrow \OIIIL[5007] emission line. The outflow is centred on the M2 UV clump in the Main component, significantly larger than the PSF (shown as a cyan-hatched region in the bottom right of the map). 

To measure the size of the outflow, we employ the curves of growth technique (COG). We extracted and fitted a spectrum of increasingly larger circular aperture centred on the peak of the outflow map with radii ranging from 0.05 to 0.4 arcseconds with steps of 0.05 arcseconds. We plot the fluxes of the outflow component normalised to the maximum flux and we plot these as red squares in the bottom panel of Figure \ref{fig:outflow_spatial}. We also repeated this analysis for an idealised NIRSpec/IFU PSF from \citet{DEugenio23ifs} and we plot the COG due to the PSF as a blue dashed line. The COG analysis shows that the outflow in the Main component of \target is spatially resolved on a scale larger than the NIRSpec/IFU PSF with a maximum radial extent of $\sim0.35$ arcsecond (deconvolved for the PSF), corresponding to a physical size of 2.1 kpc at z=6.85. 

To determine the effect of the ionised gas outflow on the gas reservoir of the galaxy we estimate the mass-loading factor $\eta$, which is defined as the ratio between the mass loss rate due to outflows (mass outflow rate) and the SFR. If $\eta$ is <1, the dominant source of gas consumption is star formation. In order to estimate the outflow rate, we assume a uniformly filled conical outflow, for which the mass outflow rate is:
\begin{equation}
\dot{M}_{\rm out}= M_{\rm out} v_{\rm out} / r_{\rm out},
\label{eq:rate}
\end{equation} 
where $M_{\rm out}$ is the mass of the outflowing gas and $r_{\rm out}$  is the extension of the outflow \citep[e.g.,][]{Maiolino12, Gonzalez-Alfonso17}. We used the standard definition of $v_{\rm out}$ from the literature of $v_{\rm out}=\Delta v \times 2\times \sigma_{\rm broad}$. The mass of the gas can be estimated from the luminosity of the broad component of \OIIIall  \citep[e.g.,][]{Carniani15}:  

\begin{equation}
M_{\rm out} = 0.8\times10^8 \left( \frac{L^{\rm corr}_{\rm [OIII]}}{\rm 10^{44}~erg~s^{-1}}\right)\left( \frac{Z_{\rm out}}{\rm Z_{\odot}}\right)^{-1}\left( \frac{n_{\rm out}}{\rm 500~cm^{-3}}\right)^{-1}~{\rm M_{\odot}},
\label{eq:moutoiii}
\end{equation} 

where $Z_{\rm out}$ and $n_{\rm out}$ are the metallicity and the electron density of the outflowing gas, respectively. The $L^{\rm corr}_{\rm H\alpha}$ and $L^{\rm corr}_{\rm [OIII]}$ are the dust corrected line luminosities of \OIIIL[5007] broad component emission lines, using the A$_{\rm V}$ from the narrow lines. As outlined in the equations above, the outflow mass is dependent on both the metallicity and electron density of the gas. Although we are able to measure the electron density from \OIIall and metallicity from \OIIIL[4363] and strong line calibrations, these quantities derived for the galaxy ISM do not necessarily match the values in the outflowing gas which tends to be more metal-enriched and for most studies at high-z are usually assumed \citep{ForsterSchreiber19,Concas19, Concas22}. Indeed stacking analysis from \citet{ForsterSchreiber19} suggests that the outflows are often denser, possibly due to compression. As we do not detect a broad component in either \OIIall or \OIIIL[4363] and we do not have wavelength coverage of [SII]$\lambda\lambda$6716,6731, we will use the values for density from \OIIall and T$_{\rm e}$ based oxygen abundance from the Main component from the bulk of the emission, which is tracing the galaxy ISM. For density and metallicity, we use the value of 1200 cm$^{-3}$ and 0.3 Z$_{\odot}$ from the T$_{e}$ method described in \S~\ref{sec:ism}.

Using the method described above we derived the mass of outflowing gas from \OIIIL[5007] to be $3.7\times 10^{6}$\Msun with an outflow mass rate of 6.1 \Msun/yr, assuming the size of 2.1 kpc, described above. However, we note that assuming a range of metallicities (0.2-0.4 Z$_{\odot}$) and densities (500--2000 cm$^{-3}$), which are values measured within different regions of the Main component and with strong line calibration, we obtain a mass outflow rate in the range of 2.7--21.5 \Msun/yr. Given the SFR of the Main component is $\sim$95 \Msun/yr (form the \Halpha), the mass loading factor is $<<1$, meaning that the gas consumption in COS-3018 is dominated by star formation rather than outflow, consistent with studies at Cosmic Noon \citep{Swinbank19, ForsterSchreiber19, Lamperti24}. 

We finally discuss the driving mechanism behind this ionised gas outflow. An outflow velocity of $\sim$1200 \kms\ is extremely rare for star-formation-driven winds, with most outflows with such velocities appearing in AGN host galaxies. It is worth pointing out that some previous works showed rare starbursts with outflow velocities in excess of 1000~\kms \citep{Diamond-Stanic12,Arribas14,Sell14,Diamond-Stanic21,Perrotta21}. However, a careful re-analysis of these objects showed that these starbursts also show signatures of AGN such as X-ray emission, \NeIV and \NeV emission lines, or lie in the composite region of the BPT diagram. As discussed in \S~\ref{sec:agn}, we do not see such definitive AGN signatures in \target, nevertheless, subtle signatures of low luminosity AGN would be hidden in the intense starburst like the one in \target. 

\subsection{Comparison with ALMA observations}\label{sec:ALMA}

In this section, we compare the ALMA \CII observations along with the rest frame optical emission lines. The full detailed presentation of the ALMA data was presented by \citet{Smit18, Witstok22, Parlanti23_alm}, here we focus on the comparison with new \JWST observations. \target was observed in both \CII and \OIIIL[88]$\mu$m. However, the \OIIIL[88]$\mu$m was observed with a compact configuration resulting in low spatial resolution (0.7 arcsec). Furthermore, the 88$\mu$m dust continuum is not detected. As a result, we focus on the comparison of JWST, \CII and 158$\mu$m dust continuum (from now on simply referred to as "dust continuum").

\begin{figure}
        \centering
	\includegraphics[width=0.9\columnwidth]{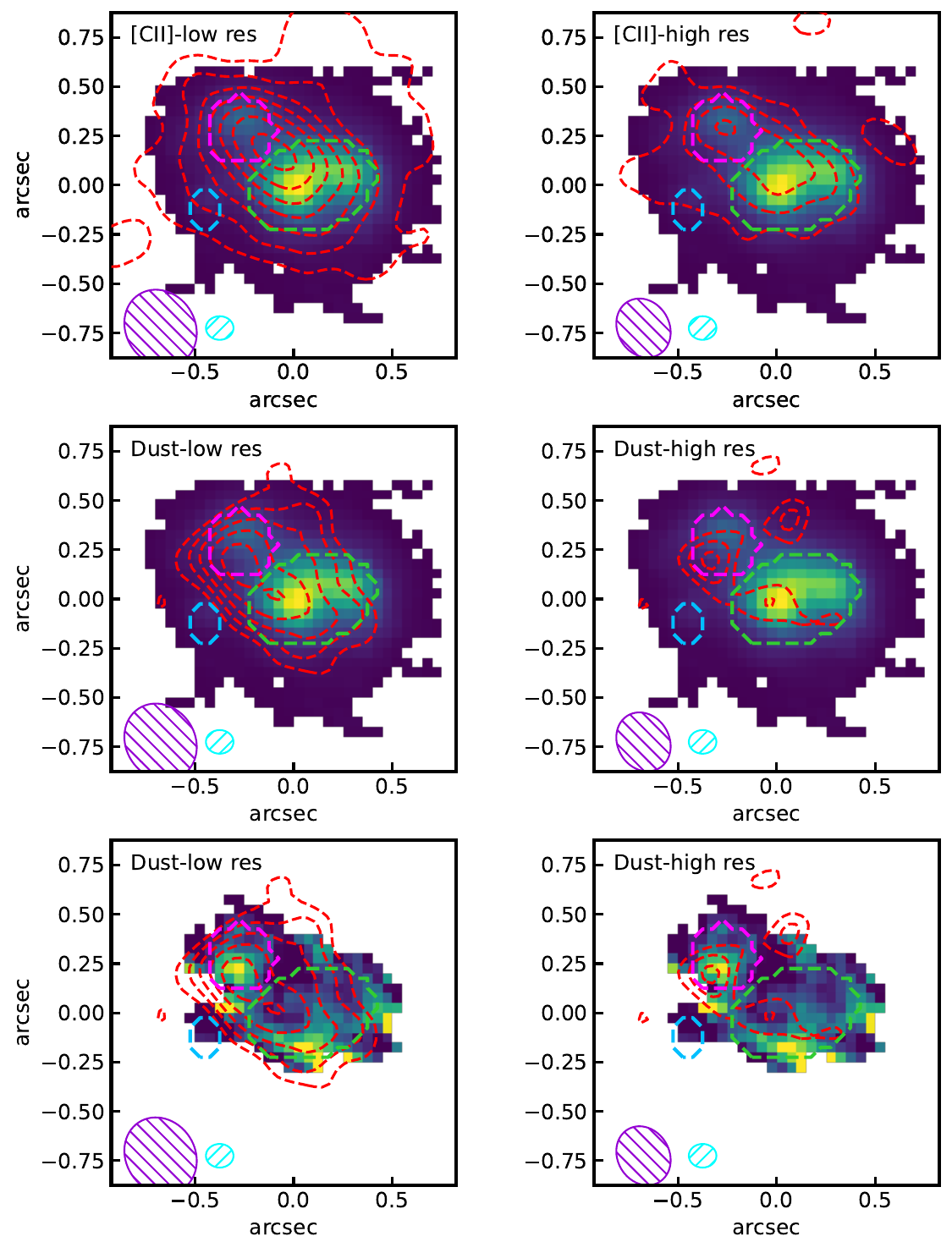}
    \caption{ Comparison of \OIIIL[5007] and A$_{\rm v}$ from NIRSpec/R2700 data and \CII and dust emission (contours from 2.5$\sigma$ with step of 1$\sigma$) from different resolutions. Left column: ALMA observations using natural tapering (low resolution). Right column: ALMA data imaged using Briggs weighting (=0.5; high resolution). From top to bottom: Comparison of \OIIIL[5007] and \CII, \OIIIL[5007] and 158$\mu$m dust continuum and A$_{\rm v}$ and 158$\mu$m dust continuum. On each map, we highlight the three separate components as in Figure \ref{fig:prism_specs}. The violet and cyan ellipses indicate the ALMA and NIRSpec PSFs.}
    \label{fig:JWST_ALMA}
\end{figure}

Leveraging on the flexibility of the ALMA observations, we imaged the \CII and dust continuum using natural and Briggs weighting (w=0.5; for more details see \S~\ref{sec:ALMAobs}) and we present these as red contours in left and right columns of Figure \ref{fig:JWST_ALMA}, respectively. We compare the ALMA observations with the \OIIIL[5007] maps (top two rows) and the derived dust attenuation (A$_{\rm v}$) map (bottom row). 

The elongated \CII emission seen in the lower resolution ALMA imaging is resolved into two separate peaks centred on the Main and North components, with the brighter \CII peak centred on the Main component. Similarly, the dust continuum is also resolved into two separate peaks centred on the Main and North components. Comparing the dust continuum and A$_{\rm v}$ map in the bottom row of Figure \ref{fig:JWST_ALMA} shows only a mild correlation between the dust map and the A$_{\rm v}$ map. While the bright dust continuum is on top of the peak of the A$_{\rm v}$ in the North component, the fainter dust continuum peak in the Main component is in a region of low A$_{\rm v}$ values. These low A$_{\rm v}$ values in the location of ALMA dust emission can be caused by rest-frame optical light from the dust-obscured region being completely obscured \citep{Chen17, Scholtz20}, resulting in anti-correlation between ALMA dust maps and dust extinction derived from optical emission lines.

The comparison of the NIRSpec and ALMA observations shows the limitations of poorly resolved emission in a complex system such as \target. \citet{Witstok22} showed a high \OIIIL[88$\mu$m]/\CII ratio, suggesting a high metallicity and ionising radiation. Despite the brightest \CII peak being on the North component, this only contains $\sim35$\% of the total \CII flux, with over 50\% of the flux located in the larger Main component. 


The increased stellar mass from the SED fitting compared to previous estimates from HST+Spitzer \citep{Bouwens15} is further relieving the tension in the large dust-to-stellar mass ratio of 0.05$^{+0.16}_{-0.04}$ reported in \citet{Witstok22}. Using the newly estimated total stellar mass of this system from \S~\ref{sec:SED_fit}, we estimate a dust-to-stellar mass ratio of 0.014$^{+0.010}_{-0.015}$. We repeat the estimate for a dust yield from supernovae and AGB stars from \citet{Witstok22} using the dust-to-stellar mass ratio to convert the dust-to-stellar mass into dust yields from \citet{Michalowski15}.

We estimated the dust yields from supernovae (y$_{\rm dust, SN}$) and AGB stars (y$_{\rm dust, AGB}$) of 1.2$^{+1.27}_{-0.9}$ \Msun and 0.41$^{+0.43}_{-0.31}$ \Msun, respectively. The estimated y$_{\rm dust, AGB}$ still exceeds the theoretical dust yield production of 0.02 \Msun even more so when the yield is increased up to two times under a different Salpeter IMF \citep[e.g.][]{Michalowski15, Lesniewska19, Schouws22}. However, the new estimate for the y$_{\rm dust, SN}$ decreases the tension with the maximum yield from supernovae of $\sim$ 1.3 \Msun \citep[][]{Todini01, Nozawa03}. However, as described in multiple works \citep{Bianchi07,  Cherchneff10, Gall11, Lakiccvic15, Ferrara21}, most of the dust produced by the supernovae is also destroyed by internal shocks drastically reducing the amount of dust by the supernovae to $<$0.1 \Msun. 
However, despite lowering the dust-to-stellar ratio by a factor of $\sim$4, our results still favour main dust production via supernovae with very low dust destruction.

\section{Conclusions}\label{sec:conclussions}

In this work, we analysed new integral field spectroscopy data from JWST/NIRSpec and imaging from JWST/NIRCam data of \target, to perform a detailed spatially resolved study of a star-burst galaxy at z$\sim$6.85. Using both high-resolution G395H grating data and the broad spectral range low resolution of the PRISM spectroscopy, we investigated the physical properties inside this star-bursting system. With this analysis, we found the following:

\begin{enumerate}
    \item \target is comprised of at least three large components, dubbed in this work as "Main", "North" and "East" components, seen across the strong emission lines and continuum in both spectroscopy and imaging. The Main and North components have a velocity offset of 300$\pm$5 \kms, while the Main and East components is 120$\pm$3 \kms.

    \item We investigated whether this system contains an AGN as hinted by previous studies. By investigating optical emission line diagnostics such as BPT, N2-He2 and \OIIIL[4363] diagnostics diagram (see Figure \ref{fig:AGN_optical}), we cannot rule out ionisation by both AGN and star formation in either of the components. The high \OIIIL[5007] equivalent width can be explained by young star-formation, however, we note that the emission line ratios are consistent with \target being type-2 AGN or a star-forming galaxy. 

    \item We used \texttt{prospector} to simultaneously model the NIRCam photometric and NIRSpec/PRISM spectroscopic data. We derived a total stellar mass of this system of $10^{9.68\pm 0.13}$ \Msun, in agreement with recent NIRCam-only estimates 0.6~dex higher than estimates from earlier \textit{HST}+\textit{Spitzer}. The SED fitting reveals a young starburst across all three components with an increase of SFR averaged over the past 10 Myr by a factor of 5-10 over the SFR averaged over the past 100 Myr.

    \item We estimated the gas-phase metallicities of the three components of \target using the strong-line calibrations and using the T$_{e}$ method using the \OIIIL[4363]. We estimated 12+log(O/H) in the range of 7.9--8.2, depending on the component and method. The gas-phase metallicities agree within the systematic uncertainties of different methods. However, we are unable to distinguish if any component is more metal-enriched than others. 

    \item We detect \NIIL[6585]in the Main and North components as well as in two regions in the spatially resolved maps, allowing us to compute N/O abundance for these regions. We estimated N/O abundance for the Main and North components of -1.2 and -1.15, respectively, consistent with what expected from its O/H abundance and from the local N/O vs O/H sequence (see Figure \ref{fig:metallicity}). We also detect \NIIL[6585] in two clumps in the Main component and west of the North component (see top right panel of Figure \ref{fig:Flux_maps}). The \NIIL[6585] clump in the Main component has $\log$ N/O abundance of -1.17, consistent with the integrated spectrum across the entire Main component. However, we find the small \NIIL[6585] clump west of the North component with $\log$ N/O = -0.8 -- this is much higher than expected for its low metallicity. This indicates that strong nitrogen enhancement can occurr in small clumps. Given the recent claims that the exceptional nitrogen enhancement may be associated with the formation of proto-globular clusters, this clump may be tracing the formation of a GC in the halo of the main galaxy.  

    \item We investigated the kinematics of the \OIIIL[5007] lines and its comparison to \CII observations (see \S \ref{sec:kinematics}). The broad velocity width of the emission lines along with the velocity profile extracted along the major and minor axes (see Figure \ref{fig:kinematics}) indicate that this system is a merger of at least three separate systems rather than a rotating disc as previously indicated by analysis of the ALMA observations.

    \item The \OIIIL[5007] emission line profile of the Main component requires three Gaussian profiles to fully describe it (see Figure \ref{sec:outflows}). The two narrower (FWHM $<$400 \kms) can be described by the kinematics of the merger component. However, we attribute the third component with FWHM of 1250 \kms\ to an outflow and measure its extent of 2.1 kpc (Figure \ref{fig:outflow_spatial}). Using the estimated size and velocity, we calculated the mass outflow rate from the \OIIIL[5007] emission line of 6.1 \Msun yr$^{-1}$ with a factor of $\sim3$ systematic uncertainties. We measure the mass loading factor (mass outflow rate/SFR) $<0.2$ showing that the dominant process consuming the gas in \target is star formation rather than outflows.

\end{enumerate}

\section*{Acknowledgements}

JS, FDE, RM and JW acknowledge support by the Science and Technology Facilities Council (STFC), ERC Advanced Grant 695671 ``QUENCH" and the
UKRI Frontier Research grant RISEandFALL.
RM also acknowledges funding from a research professorship from the Royal Society.
S.C, EP and GV acknowledge support from the European Union (ERC, WINGS,101040227). 
H\"U acknowledges support through the ERC Starting Grant 101164796 ``APEX''.
IL acknowledges support from PRIN-MUR project "PROMETEUS" (202223XPZM).
MP, SA and BRdP acknowledge grant PID2021-127718NB-I00 funded by the Spanish Ministry of Science and Innovation/State Agency of Research (MICIN/AEI/ 10.13039/501100011033)
PGP-G acknowledges support from Spanish Ministerio  de  Ciencia e Innovaci\'on MCIN/AEI/10.13039/501100011033 through grant PGC2018-093499-B-I00. The work of CCW is supported by NOIRLab, which is managed by the Association of Universities for Research in Astronomy (AURA) under a cooperative agreement with the National Science Foundation. 
AJB and GCJ acknowledge funding from the "FirstGalaxies" Advanced Grant from the European Research Council (ERC) under the European Union’s Horizon 2020 research and innovation programme (Grant agreement No. 789056).
GC acknowledges the support of the INAF Large Grant 2022 ""The metal circle: a new sharp view of the baryon
cycle up to Cosmic Dawn with the latest generation IFU facilities""
SA acknowledges support from the JWST Mid-Infrared Instrument (MIRI) Science Team Lead, grant 80NSSC18K0555, from NASA Goddard Space Flight Center to the University of Arizona.

\section*{Data Availability}

The datasets were derived from sources in the public domain: ALMA data from \url{https://almascience.nrao.edu/aq/?result_view=observation} and JWST/NIRSpec IFS data from MAST/



\bibliographystyle{mnras}
\bibliography{mybib,mirko_abudances} 




\appendix

\section{Regional spectra}\label{sec:spec_reg_full}

\begin{figure*}
        \centering
	\includegraphics[width=0.9\paperwidth]{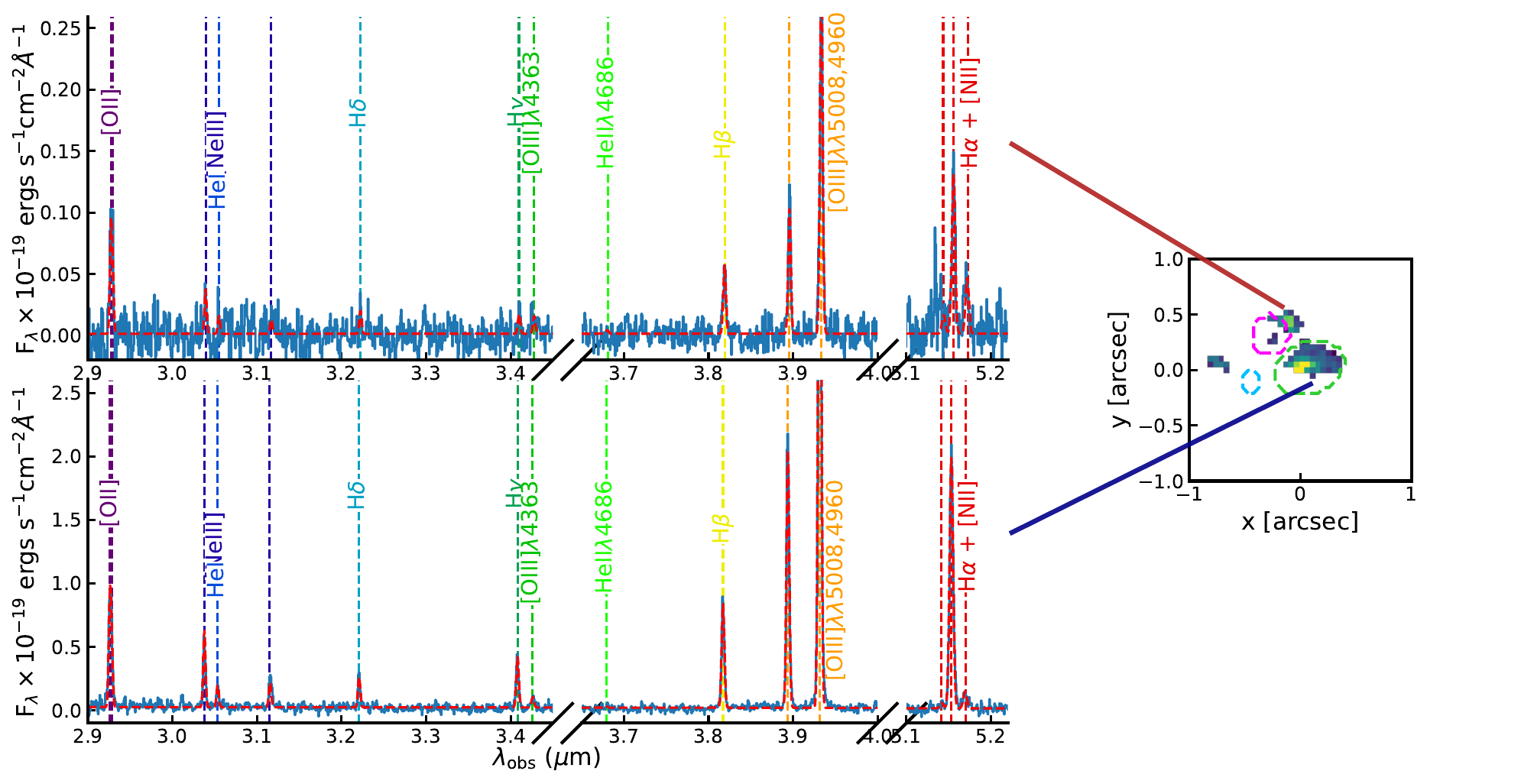}
    \caption{Top left: Spectrum North \NII clump component. Bottom left:  Spectrum of the \NII clump in the Main component. Right panel: Map of the \NIIall emission with the coloured contours indicating the Main, North and East components of \target.}
    \label{fig:NII_clumps}
\end{figure*}

In figure \ref{fig:NII_clumps}, we present the R2700 spectra of the two \NII clumps in the Main and North components.
 
\section{Prospector fits}\label{sec:prosp_fits_full}

\begin{figure*}
        \centering
	\includegraphics[width=0.9\paperwidth]{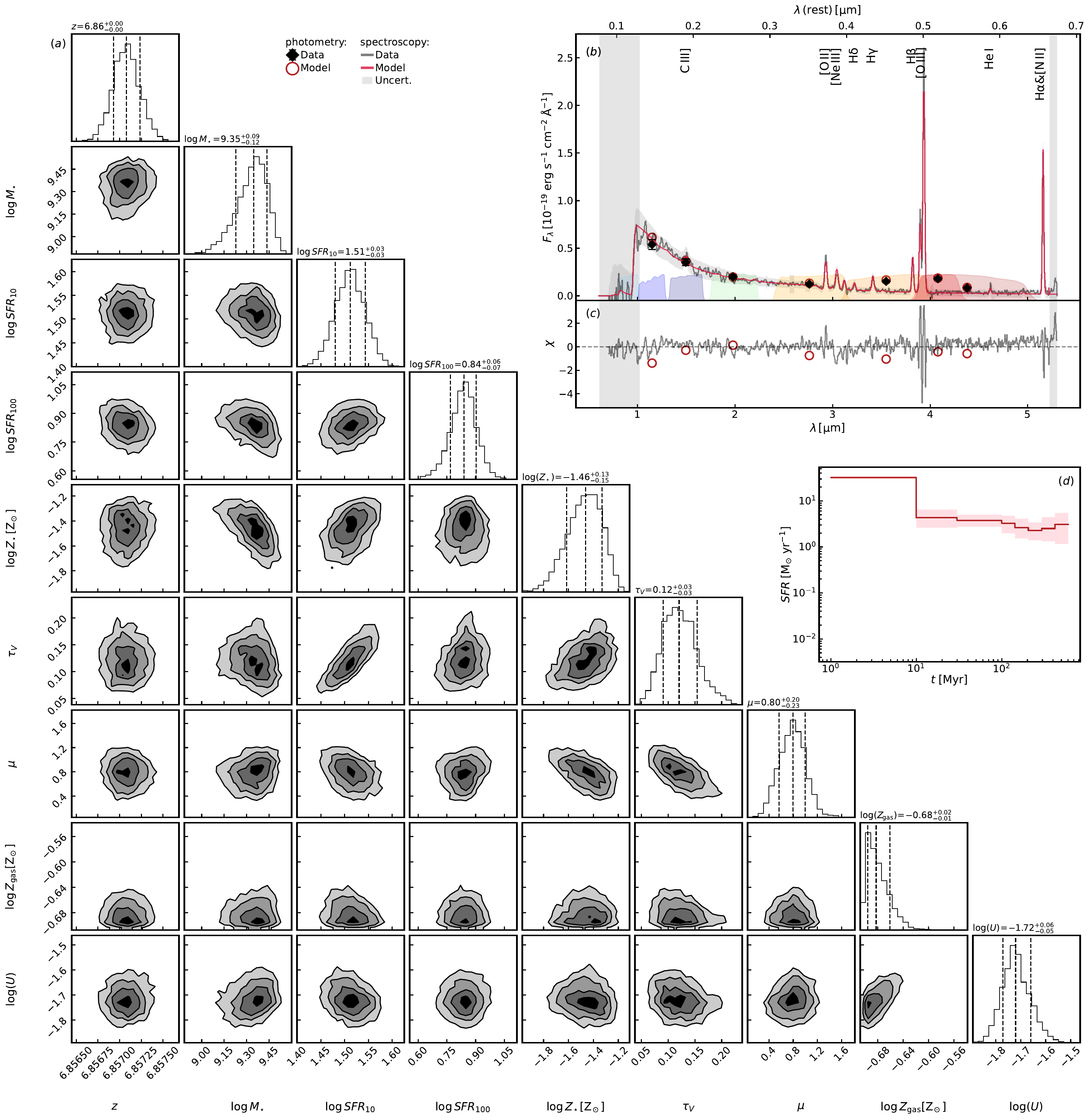}
    \caption{\texttt{Prospector} fitting results of the Main component. a) Corner plot of the fitted parameters: redshift, stellar mass, SFR averaged over 10 Myr, SFR averaged over 100 MyR, stellar metallicity, optical depth of the diffuse ISM, extra dust attenuation towards birth clouds, gas metallicity, ionisation parameter. b) Data (black) and the best fit model (red) for the photometry and spectroscopic data. c) visualisation of the residuals of the photometric and spectroscopic data. d) Star-formation history with the red shaded region indicating the 68 \% confidence interval. }
    \label{fig:prosp_main}
\end{figure*}

\begin{figure*}
        \centering
	\includegraphics[width=0.9\paperwidth]{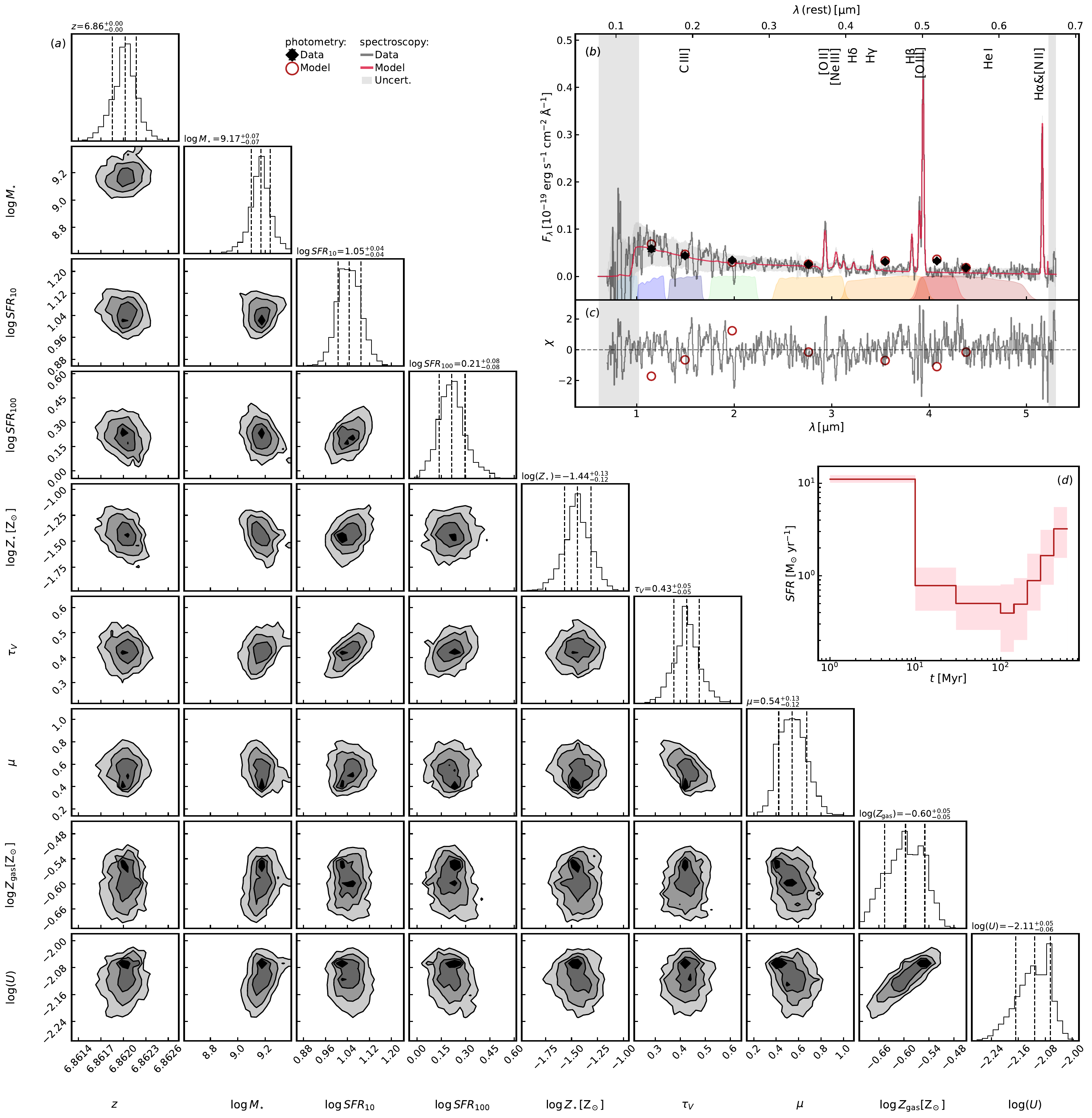}
    \caption{Same as Figure \ref{fig:prosp_main}, but for the North component. }
    \label{fig:prosp_north}
\end{figure*}

\begin{figure*}
        \centering
	\includegraphics[width=0.9\paperwidth]{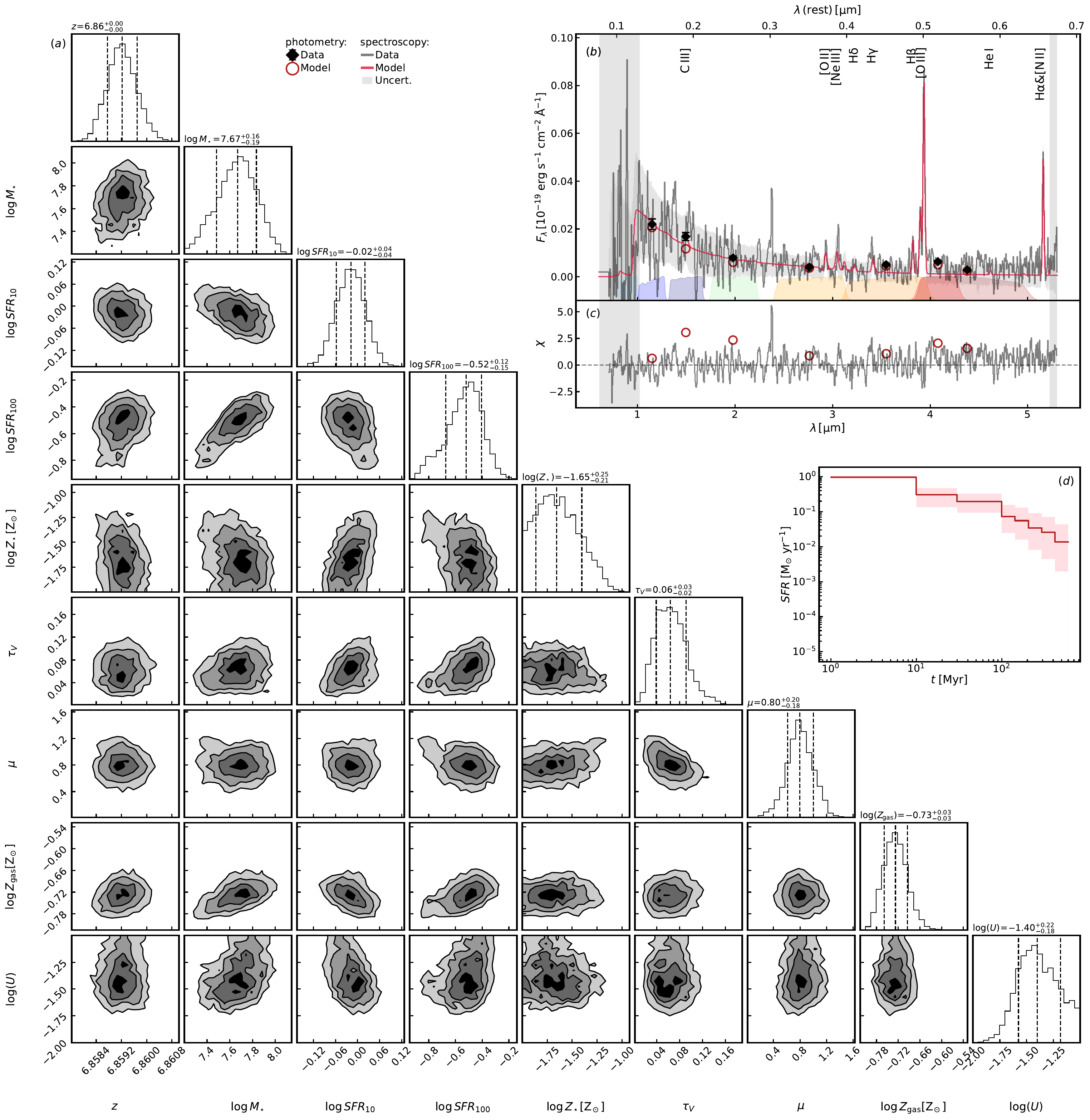}
    \caption{Same as Figure \ref{fig:prosp_main}, but for the East component.}
    \label{fig:prosp_east}
\end{figure*}

 In Figures \ref{fig:prosp_main}, \ref{fig:prosp_north}, \ref{fig:prosp_east} we show the full results of the \texttt{prospector} fitting described in \S~\ref{sec:SED_fit}.
 
\bsp	
\label{lastpage}
\end{document}

%% file: Table_fluxes.tex
\begin{tabular}{lcccccc} 
\hline 
\hline 
Object & \multicolumn{2}{c}{Main} & \multicolumn{2}{c}{North} & \multicolumn{2}{c}{East} \\
 line &  F$_{\rm PRISM}$ & F$_{\rm F290LP}$ & F$_{\rm PRISM}$ & F$_{\rm F290LP}$ & F$_{\rm PRISM}$ & F$_{\rm F290LP}$ \\
\hline 
\Halpha+\NIIall& 272.5$\pm$5.9& 248.0$\pm$ 1.7& 56.6$\pm$1.3& 58.3$\pm$ 0.9& 6.1$\pm$0.4& 4.3$\pm$ 0.2\\
\Halpha& -& 234.4$\pm$1.9& -& 51.5$\pm$ 0.9& -& 3.9$\pm$ 0.2\\
\NIIall& -& 14.2$\pm$1.1& -& 5.1$\pm$ 0.8& -&<0.8\\
\OIIIall& 528.7$\pm$6.2& 538.5$\pm$2.2& 88.6$\pm$1.1& 90.9$\pm$ 0.8& 12.2$\pm$0.3& 8.6$\pm$ 0.2\\
\Hbeta& 76.6$\pm$3.2& 65.9$\pm$1.0& 14.1$\pm$0.9& 14.6$\pm$ 0.4& 2.4$\pm$0.3& 1.4$\pm$ 0.2\\
\HeIIL[4686]& <8.7&<2.4& <2.3&<0.4& <0.8&<0.9\\
H$\gamma$& 27.5$\pm$3.2& 29.2$\pm$1.0& 6.5$\pm$0.8& 6.5$\pm$ 0.4& <0.9& 0.6$\pm$ 0.1\\
\OIIIL[4363]& 10.8$\pm$3.0& 7.0$\pm$0.9& <2.2& 2.0$\pm$ 0.4& <0.9&<1.5\\
H$\delta$& 15.3$\pm$2.9& 14.9$\pm$1.8& 4.4$\pm$0.8& 3.5$\pm$ 0.6& <0.9&<1.1\\
\NeIIIL& 35.7$\pm$3.3& 36.4$\pm$1.2& 6.6$\pm$0.9& 6.4$\pm$ 0.5& <1.1&<1.6\\
\OIIall& 81.3$\pm$3.7& 85.7$\pm$2.7& 26.1$\pm$1.0& 26.8$\pm$1.2& 2.4$\pm$0.4& <2.5\\
\OIIL[3727]& -& 48.4$\pm$2.1& -& 13.9$\pm$ 0.7& -&<1.4\\
\OIIL[3729]& -& 37.4$\pm$2.0& -& 12.9$\pm$ 0.8& -&<1.1\\
\hline 
CIII]$\lambda$1906 & <23.1& - & <29.7& - & <16.8 & - \\
NIII]$\lambda$1750 & <36.1& - & <33.1& - & <19.5 & - \\
\HeIIL[1640]+ \OIIIL[1660] & <28.8 & - &<39.0 & - & <22.8& - \\
CIV]$\lambda$1550 & <33& - &<42.1 & - & <24.9& - \\
NIV]$\lambda$1497 & <26.1& - & <45.0& - & <27.6 & - \\
\hline
\hline
\end{tabular} 